\documentclass[12pt]{iopart}
\pdfoutput=1
\usepackage{color}
\usepackage{amssymb}

\usepackage{graphicx} 

\def\beqra{\begin{eqnarray}}
\def\eeqra{\end{eqnarray}}
\def\beq{\begin{equation}}
\def\eeq{\end{equation}}
\def\etap{\eta^\prime}

\def\ds{\displaystyle}

\def\vp{\varphi}

\def\bx{{\bf{x}}}

\def\bk{{\bf{k}}}
\def\bp{{\bf{p}}}
\def\bq{{\bf{q}}}

\def\bv{{\bf{v}}}
\def\bw{{\bf{w}}}

\def\bV0{{\bf{V_0}}}
\def\re#1{(\ref{#1})}

\begin{document}

\begin{flushright}
{\small UMN-TH-3310/13}
\end{flushright}
\title{
Ward identities and consistency relations for the large scale structure with multiple species 
}
\author{Marco Peloso$^{1}$}
\ead{peloso@physics.umn.edu}
\author{Massimo Pietroni$^{2,3}$}
\ead{pietroni@pd.infn.it}
\vskip 0.3 cm
\address{
$^1$School of Physics and Astronomy, University of Minnesota, Minneapolis, 55455, USA\\
$^2$INFN, Sezione di Padova, via Marzolo 8, I-35131, Padova, Italy\\
$^3$Dipartimento di Fisica e Scienze della Terra, Universit\`a di Parma, Viale Usberti 7/A, I-43100 Parma, Italy
}

\begin{abstract}

We present fully nonlinear consistency relations for  the squeezed bispectrum of  
 Large Scale Structure. These relations hold when the matter component of the Universe is composed of one or more  species, and generalize those obtained in  \cite{Peloso:2013zw,Kehagias:2013yd} in the single species case. The multi-species relations apply  to the standard dark matter $+$  baryons scenario, as well as to the case in which some of the fields are auxiliary quantities describing a particular population, such as dark matter halos or a specific galaxy class.  
 
 If a large scale velocity bias exists between the different populations new terms appear in the consistency relations with respect to the single species case. As an illustration, we discuss two physical cases in which such a velocity bias can exist: (1) a new long range scalar force in the dark matter sector (resulting in a violation of the equivalence principle in the dark matter-baryon system), and (2) the distribution of dark matter halos relative to that of the underlying dark matter field.

\end{abstract}

\maketitle

\section{Introduction}

The role of the Large Scale Structure (LSS) of the Universe as the main source of information on the latest stages of the evolution of the Universe is widely recognized, and it will be extensively exploited by future surveys. On the theory side, the complementarity between N-body simulations and semi-analytical methods is emerging as the optimal strategy to compute the relevant observables in the widest possible class of cosmological models. It is well known that cosmological perturbation theory (PT)  \cite{PT} fails at low redshifts and small scales, while N-body simulations of the required accuracy are too expensive in terms of computational times to make a thorough exploration of parameters/models spaces feasible. For these reasons, semi-analytical methods based on reorganizations of the perturbative expansion 
\cite{RPTa,RPTb,MP07b, RPTBAO, Taruya2007, Matsubara07, Pietroni08, BV08, Bernardeau:2008fa, Anselmi:2010fs, Wang:2011fj, Wang:2012fr}, or hybrid approaches in which the small scales information is provided by N-body simulations while the large scales one is computed (semi-)analytically \cite{Pietroni:2011iz,Carrasco:2012cv}, have been recently investigated  (see also \cite{Hertzberg:2012qn}). 

Although any practical computation requires some approximation, gaining information on the structure of the full ({\it i.e. not approximated}) result is of utmost importance, both as a way to test cosmological models vs. observations, and as a consistency check of different approximation schemes. An example of such non-perturbative information is provided by Ward identities (WI) and consistency relations (CR), which have been recently derived in the LSS context in \cite{Peloso:2013zw,Kehagias:2013yd} exploiting the Galilean invariance of the underlying newtonian theory. Later, ref.~\cite{Creminelli:2013mca} derived  CR for the density contrast correlators starting from a relativistic description of the dark matter system. The relations of 
\cite{Creminelli:2013mca} follow from diffeomorfism invariance, and reduce to those of  \cite{Peloso:2013zw,Kehagias:2013yd} in the non-relativistic regime\footnote{See also refs.~\cite{Hinterbichler:2013dpa,Berezhiani:2013ewa} for derivations of CR for cosmological correlators coming from WI.}. 

In this paper, we extend the derivation of \cite{Peloso:2013zw} to the case in which the matter content of the Universe is made up of more than one species. This applies not only to the baryon-dark matter (DM) case of the presently favoured $\Lambda$CDM model, but also to more generic scenarios, such as those in which the dark sector has different components, possibly with a new long-range force coupled differently to them. Moreover, this formalism can be applied also to the study of halo bias, as done in \cite{2011MNRAS.416.1703E}. 

The consideration of multi-species clarifies the relation between WI and CR in the LSS context. Namely, as in quantum field theory, WI are the direct consequence of the symmetries of the equations of motion, or, equivalently, of the action. These relations provide constraints between the different terms of the full action, which have to be satisfied in any viable approximation scheme. To pass from WI to the CR between correlators, one has to add information on the structure of the solutions of the equations of motion at large scales. 
At the level of the equations of motion, the symmetry under consideration is a generalized Galilean invariance ({\it i.e.} invariance by time-dependent boosts)  both in the single-species and in the multi-species case. To get the consistency relation between the bispectrum and the power spectrum (PS) in the single-species case we use the fact that a long wavelength velocity mode, solution of the equations of motion, can be reabsorbed by a Galilean transformation. This may not be the case in the multi-species case if velocity bias exists among the different species. In this case, new terms appear in the CR in addition to those expected from a straightforward generalization of the single species case.

 It is however noticeable that, apart from an expansion in $(1-b_v)$, where $b_v$ is the large scale velocity bias, even these new terms are exact nonperturbatively, and may be used as a tool to measure velocity bias in observations.  The situation is somehow analogous to what happens in quantum field theory with spontaneous symmetry breaking, where one can write WI and exact relations between Green's functions even in the broken phase, where the solutions of the field equations do not respect the symmetry of the action.

Our main result is given by eq. \re{nexwardbias}, which is an expression for the squeezed (self- or coss-) three point correlation function  of density and / or velocity fields of two  species  at arbitrary times. While the single species consistency relation obtained in  \cite{Peloso:2013zw,Kehagias:2013yd} vanishes at equal times, this is not the case for the multi-species relation that we obtain here. This could facilitate an observational test of this relation. The expression for the equal time ($\tau'' = \tau'$)  cross-correlation of the density contrast $\delta_a$  of the species $1$ and $2$, reads

\begin{eqnarray} 
& & \!\!\!\!\!\!\!\!  \!\!\!\!\!\!\!\!  \!\!\!\!\!\!\!\!  \!\!\!\!\!\!\!\!   \!\!\!\!\!\!\!\!  
\lim_{k \ll q } B_{112}^{(\delta)} \left( k, q, \vert \bk + \bq \vert ; \tau, \tau' , \tau' \right) = 
  - P_{11,L}^{\left(\delta v\right)} \left( k; \tau, \tau \right) \left[ 1 - b_v \left( k ,\tau\right) \right]  \frac{\bk \cdot \bq}{k^2} 
\nonumber\\ 
& &     
 \left\{ \frac{D_+ \left( \tau' \right)}{D_+ \left( \tau \right)} P_{12}^{(\delta)} \left( q; \tau', \tau' \right) 
  -   \int^{\tau'}_{\tau_{in}} d {\tilde \tau} \; 
{\cal H}^2 \left( {\tilde \tau} \right) f^2 \left( {\tilde \tau} \right) 
 \frac{D_+ \left( {\tilde \tau} \right)}{D_+ \left( \tau \right)} 
  {\cal T} \left( q; {\tilde \tau}, \tau' \right) 
   \right\} \nonumber\\ 
& & \!\!\!\!\!\!\!\!  \!\!\!\!\!\!\!\!  \!\!\!\!\!\!\!\!  \!\!\!\!\!\!\!\!      + {\rm O } \left( k^0 ; \left( 1 - b_v(k,\tau) \right)^2 \right)  \;, 
\label{consistency-delta}
\end{eqnarray} 
where
\begin{eqnarray}
& & \!\!\!\!\!\!\!\! \!\!\!\!\!\!\!\! \!\!\!\!\!\!\!\!  \!\!\!\!\!\!\!\! 
{\cal T} \left( q; {\tilde \tau}, \tau' \right) \equiv 
 \int d^3p'\,\frac{\bp'\cdot\bq}{q^2}  \Bigg[ \Omega_{\theta_1,\delta_2}   \left\langle \frac{\partial \left( \delta_2 \left( \bp' , {\tilde \tau} \right)  \delta_1 \left( \bq , \tau' \right)   \delta_2 \left( - \bq , \tau' \right) \right) }{ \partial \, \theta_1 \left( \bp', {\tilde \tau} \right) } \right\rangle' \nonumber\\
 & & \qquad\qquad \qquad \qquad \quad -  \Omega_{\theta_2,\delta_1}   \left\langle \frac{\partial \left( \delta_1 \left( \bp' , {\tilde \tau} \right)  \delta_1 \left( \bq , \tau' \right)   \delta_2 \left( - \bq , \tau' \right) \right) }{ \partial \, \theta_2 \left( \bp', {\tilde \tau} \right) } \right\rangle'   \Bigg] \;, 
\label{T-intro}
\end{eqnarray}
and where  the power spectra and bispectrum are defined in eqs.  \re{bispa},  \re{PS}, and  \re{P1beta}, 
the large scale velocity bias $ b_v(k,\tau)$  is defined in eq. \re{P1beta} - or, equivalently, in eq. \re{d3} -  $\theta_i$ is the divergence of the velocity field of the species $i$, ${\cal H}\left( \tau \right)$ is the Hubble rate in conformal time $\tau$,  $D_+$ is the linear growth factor, $f = d \log D_+ / d \log a$ (with $a$ indicating the scale factor). The elements $\Omega_{\theta_1,\delta_2}$ and $\Omega_{\theta_2,\delta_1}$ of the matrix $\Omega$, introduced in eq.  \re{BIGOMEGA}, give the coupling between the two species.~\footnote{The indices  in eq. \re{consistency-delta} indicate the species ($1$ or $2$) to which each quantity refers. This is the convention used in Section  \ref{gaussian}. When instead in the remainder of the paper we consider a system of two particles, the indices $1$ and $2$ refer, respectively, to the density contrast and velocity divergency of the first species (rescaled as in eq. \re{multiplet}), while the  indices $3$ and $4$ refer, respectively, to the density contrast and velocity divergency of the first species (rescaled as in eq. \re{multiplet}). For this reason,  the two coefficients  $ \Omega_{\theta_1,\delta_2} $ and $  \Omega_{\theta_2,\delta_1}  $ are denoted, respectively, as $\Omega_{23}$ and $\Omega_{41}$ in eq.  \re{gil}.} 

The first term in the second line of  eq. \re{consistency-delta} is a kinematic contribution that arises because the two species have a different large scale velocity, while the other term  is    due to the coupling between the two species via the gravitational interaction and, possibly, via an extra long range interaction . The symbol $\partial$ in the expectation values in \re{T-intro}  denote functional differentiation, while prime indicates that an overall Dirac delta function $\delta_D$ has been factored out from them. 
 
This result can apply to  various physical situations. After a general treatment, we specialize to two special cases. One is the already mentioned case in which two matter species exist coupled to a long range scalar field with different strength, thus violating the equivalence principle (EP). A velocity bias exists in this case already at the linear level, which is directly computable in terms of the new couplings. 
In our second example, we consider the clustering properties of DM halos with respect to the full DM distribution. In this case, a velocity bias exists between the two populations as a statistical selection effect on the initial conditions for the two fluids due to the fact that, roughly speaking, halos are localized about the peaks of the DM distribution \cite{Desjacques:2009kt,Elia:2011ds}. The CR we derive in this case can be relevant, for instance, when cross-correlating galaxy surveys with weak lensing ones.~\footnote{ A third example, that we do not consider here, is a velocity bias arising from  isocurvature modes in the  baryon-DM fluid \cite{Tseliakhovich:2010bj}.}

From an observational point of view, the CR derived in the single species case \cite{Peloso:2013zw,Kehagias:2013yd} are quite challenging, as  they involve correlators (PS and bispectrum) between fields at different cosmic times. While these objects are easily computed in a N-body simulations by cross-correlating different snapshots, they can hardly be tested in the short scale regime in actual observations, which are confined on the light cone. On the other hand, as  eq. \re{consistency-delta} shows,  in the multi-species case the new terms induced by velocity bias do not vanish for equal time  correlators, and could therefore provide interesting tests of the velocity fields. However, it has to be emphasized that, while a squeezed bispectrum $ B_{112}^\delta \left( k, q, \vert \bk + \bq \vert ; \tau, \tau' , \tau' \right)$ proportional to $\bq\cdot\bk$ would indicate a nonvanishing velocity bias, also the CR \re{consistency-delta} is hard to test observationally, due to the $\langle \frac{\partial \vp^3 }{\partial \vp} \rangle$ terms. Analogously to the propagator (see. eq. \re{G-def2}), such terms indicate the response of the $3-$point correlation function to a change of the value of the field (in this case, the velocity one) at some previous time, and therefore they are not easily accessible to observations done on the light cone, particularly for nonlinear modes.  \footnote{Two points separated by  $\Delta z$ in the radial direction are separated by $\Delta z/H \sim \Delta z \;  3000 {\rm Mpc}/h$. Probing nonlinear scales therefore requires cross-correlating different redshift slices whose average redshifts differ by $\Delta z < O(10^{-2})$.}  Most likely, these  terms need to be measured in N-body simulations, or computed perturbatively (see eq. \re{nexwardbias2}).

The paper is organized as follows: in Sect.~\ref{gaussian} we discuss the emergence of the CR  as the effect of non-trivial translations in position space, in Sect.~\ref{invariance-tree} we introduce our formalism and discuss the invariance under generalized Galilean transformations of the LSS equations of motion.  In Sect.~\ref{ward} we study the WI associated with these transformations. In Sect.~\ref{consistency} we discuss how one can obtain  CR between the squeezed bispectrum and the power spectrum. These relations are valid at the full nonlinear level, and generalize those obtained in  \cite{Peloso:2013zw} to the case of two species with a velocity bias. In Sect.~\ref{examples} we briefly discuss  two cases in which a nontrivial velocity bias can arise: namely from violations of the EP, and the halo bias.  In Section \ref{bssa} we show that the CR   are not modified by multi-streaming and by the standard photon-baryon interactions.  In Sect.~\ref{conclus} we present our conclusions. In \ref{app:WICR} we show how the CR can be explicitly obtained from the WI. In ~\ref{treetest} we provide an explicit tree level check of the consistency relation, and in \ref{isorole} we study the role of the isocruvature mode in the equal time squeezed bispectrum, where we show that  the equal time squeezed bispectrum vanishes in absence of these modes, in agreement with what obtained in  \cite{Peloso:2013zw}. Finally, in \ref{appA}, we derive some useful properties of the propagator.

\section{Large scale motions and consistency relations}
\label{gaussian}

The physical content of the CR derived in  \cite{Peloso:2013zw,Kehagias:2013yd} and of those to be discussed in this paper is the response of the correlators of the density and velocity fields to a large scale motion. In particular, in this work we study how the presence of multi-species with a nontrivial velocity bias modifies the relations obtained in  \cite{Peloso:2013zw,Kehagias:2013yd} in the case of a single species. There are two distinct modifications:  (i) the first is a purely kinematic effect, caused by the different large scale motion of the two species; (ii) the second is due to the coupling (gravitational, and, possibly, from  additional long-distance forces) between  the different species. Both effects are consistently included in the computations that we perform in the following sections, and that lead to our final result  (\ref{nexwardbias}). In this Section, we instead  obtain an exact expression for the squeezed bi-spectrum of different species induced by the effect (i), in the (unphysical) limit in which the effect (ii) is neglected. We do so because the effect (i) can be intuitively understood, and accounted for, following the same lines leading to the CR of  \cite{Peloso:2013zw,Kehagias:2013yd} and, therefore, this derivation clarifies the purey kinematical origin of such CR. The result that we obtain in this section can be immediately identified  with the second and third line   of the  full equation  (\ref{nexwardbias}). 

We start by considering the continuity equation for a species $\alpha$
\beq
\frac{\partial}{\partial \tau} \delta_\alpha (\bx,\tau) +\frac{\partial}{\partial x^i} [(1+\delta_\alpha (\bx,\tau))v_\alpha^i(\bx,\tau)]=0\,,
\eeq 
and split the velocity field in a short-scale and a long-scale component
\beq
\bv_\alpha(\bx,\tau)=\bv_{\alpha,{\rm short}}(\bx,\tau)+\bv_{\alpha,{\rm long}}(\bx,\tau)\,, 
\label{splitV}
\eeq
so that the continuity equation becomes
\beq
\frac{\partial}{\partial \tau} \delta_\alpha (\bx,\tau) +\frac{\partial}{\partial x^i} [(1+\delta_\alpha (\bx,\tau))v^i_{\alpha,{\rm short}}(\bx,\tau)] + v^i_{\alpha,{\rm long}}(\bx,\tau)\frac{\partial}{\partial x^i}\delta_\alpha (\bx,\tau) \simeq 0\,,
\label{conti}
\eeq
where we have neglected a term containing the gradient of $v^i_{\alpha,{\rm long}}(\bx,\tau)$. 
We define the shifted field
\beq
\bar\delta_\alpha(\bx,\tau) = \delta_\alpha(\bx -{\bf D}_\alpha (\bx,\tau),\tau),
\label{filrep}
\eeq
with 
\beq
{\bf D}_\alpha (\bx,\tau) \equiv \int_{\tau_{in}}^\tau d\tau' \bv_{\alpha,{\rm long}}(\bx,\tau') \,.
\label{dispr}
\eeq
In terms of ${\bar \delta}_\alpha$, eq.~\re{conti} rewrites
\beq
\frac{\partial}{\partial \tau} \bar\delta_\alpha (\bx,\tau) +\frac{\partial}{\partial x^i} [(1+\bar\delta_\alpha(\bx,\tau))v^i_{\alpha,short}(\bx,\tau)] \simeq 0\,,
\eeq
that is, the effect of the long-wavelength field (neglecting its gradient) is reabsorbed by shifting the space coordinate of the original field, and disappears from the non-linear (mode-coupling) terms.

If there is a single species, or if there are different species with equal long-wavelength velocity field, then it is immediate to verify that this shift is also a symmetry of the full theory (namely, that shifting all the density and velocity fields as in \re{filrep} removes the long wavelength velocity fields also from the Euler and Poisson equations). We can therefore express correlators in terms of the $\delta_\alpha$ or, equivalently, the ${\bar \delta}_\alpha$ fields. For multiple species with  nontrivial velocity bias ($b_v \neq 1$) this would be the case only if the effect (ii) mentioned above is disregarded. As we already discussed, we do so in the present Section. 

We consider the bispectrum
\beqra
&&\!\!\!\!\!\! \!\!\!\!\!\!\ \!\!\!\!\!\!\!\!\!\!\!\! \!\!\!\!\!\!\!\!\!\!\!\!  B^{(\delta)}_{\alpha\beta\gamma}(\bk,\bp,\bq;\tau,\tau',\tau'') \delta_D \left( \bk + \bp + \bq \right)= \nonumber\\
&&\!\!\!\!\!\!\ \!\!\!\!\!\!\ \int \frac{d^3 x_1}{(2\pi)^3} \frac{d^3 x_2}{(2\pi)^3} \frac{d^3 x_3}{(2\pi)^3} e^{-i(\bk\cdot(\bx_1-\bx_2) -\bq\cdot(\bx_2-\bx_3))}\langle\delta_\alpha(\bx_1,\tau)\delta_\beta (\bx_2,\tau')\delta_\gamma (\bx_3,\tau'') \rangle\nonumber\\
&&\!\!\!\!\!\!\ \!\!\!\!\!\!\ = \int \frac{d^3 x_1}{(2\pi)^3} \frac{d^3 x_2}{(2\pi)^3} \frac{d^3 x_3}{(2\pi)^3}  e^{-i(\bk\cdot(\bx_1-\bx_2) -\bq\cdot(\bx_2-\bx_3))}\nonumber\\
&&\times \langle \bar\delta_\alpha(\bx_1+{\bf D}_\alpha(\bx_1,\tau),\tau)\bar\delta_\beta(\bx_2+{\bf D_\beta}(\bx_2,\tau'),\tau')\bar\delta_\gamma(\bx_3+{\bf D_\gamma}(\bx_3,\tau''),\tau'')\rangle\,.\nonumber\\
\label{bispa}
\eeqra
Taylor-expanding in the displacement fields ${\bf D}$, we obtain  
\beqra
&& \!\!\!\!\!\!\!\!\!\!\!\! \!\!\!\!\!\!\!\!\!\!\!\! \!\!\!\!\!\!\!\!\!\!\!\!  \langle \bar\delta_\alpha(\bx_1+{\bf D}_\alpha(\bx_1,\tau),\tau)\bar\delta_\beta(\bx_2+{\bf D}_\beta(\bx_2,\tau'),\tau')\bar\delta_\gamma(\bx_3+{\bf D}_\gamma(\bx_3,\tau''),\tau'')\rangle =  \nonumber\\
&&\!\!\!\!\!\!\!\!\!\!\!\!   \langle \bar\delta_\alpha(\bx_1,\tau)\bar\delta_\beta(\bx_2,\tau')\bar\delta_\gamma(\bx_3,\tau'')\rangle+\langle  d_\alpha^i(\bx_1,\tau)\frac{\partial}{\partial x_1^i} \bar\delta_\alpha(\bx_1,\tau)\bar\delta_\beta(\bx_2,\tau')\bar\delta_\gamma(\bx_3,\tau'')\rangle\nonumber\\
&& + \;\;2\;\;{\mathrm{ cyclic\;\; permutations}} + {\rm \; higher \; orders }  + \cdots\,,
  \label{expa} 
\eeqra 
where the dots indicate extra contributions in ${\bf D}$, which would be present if the effect (ii) would be taken into account  (see Sect.~\ref{consistency}). This  extra ${\bf D}$-dependence arises in the Poisson equation, and in the Poisson-like equations for  extra long-range scalar fields \cite{Am03, Saracco:2009df}, when shifts  of the type \re{filrep} act differently on different species. On the other hand, in the following, we will  include all the higher orders of the Taylor expansion (\ref{expa}). 

 In the following, the symbol $ \cong $ indicates  equality up to the effect (ii). We are interested in the squeezed configuration for the bispectrum, namely
\beq
k\ll q \simeq p \,,
\label{sql}
\eeq
where in the triangle identifying the bispectrum in Fourier space the two sides carrying the hard momentum $\bq  \simeq - \bp $ correspond to different times $\tau'$ and $\tau''$\footnote{In real space it corresponds to having $|\bx_1-\bx_2|\gg |\bx_2-\bx_3|$.}, and, moreover, we consider long velocity modes which are nearly constant on the short scale $\sim 1/q\sim |\bx_2-\bx_3|$, that is~\footnote{To avoid having a too heavy notation, we denote with the same symbol a field and its Fourier transform.} 

\beq
 {\bf D}_{\alpha,\beta,\gamma}({\bf p'},\tau)\neq 0, \;\;\;\mathrm{only \;\;for}\;\;\;p'\ll q\,.
\label{displ}
\eeq 
In the regime \re{sql}, we can disregard the difference between the long wavelength fields $\delta_\alpha \left( \bk \right)$ and  ${\bar \delta}_\alpha \left( \bk \right)$  and substitute $\bp$ with $-\bq$ in the momentum emerging from derivatives of ${\bar \delta}_\beta$. It is actually convenient to go in a frame where $ {\bf D}_\beta \rightarrow 0$ and 
 $ {\bf D}_\gamma \rightarrow \Delta  {\bf D}_{\gamma\beta}$, defined as 
\begin{equation}
\Delta  {\bf D}_{\gamma\beta}(\bk;\tau'',\tau') \equiv  {\bf D}_\gamma (\bk,\tau'')- {\bf D}_\beta (\bk,\tau')\,. 
\label{reldisp}
\end{equation}
We then obtain, considering all higher orders in the expansion in eq.~\re{expa}
\beqra
&&\!\!\!\!\!\! \!\!\!\!\!\!\ \!\!\!\!\!\!\!\!\!\!\!\! \!\!\!\!\!\!\!\!\!\!\!\!  B^{(\delta)}_{\alpha\beta\gamma}(\bk,\bp,\bq;\tau,\tau',\tau'') \delta_D \left( \bk + \bp + \bq \right) \nonumber\\
& & \!\!\!\!\!\! \!\!\!\!\!\!\ \!\!\!\!\!\!\!\!\!\!\!\! \!\!\!\!\!\!\!\!\!\!\!\!   \cong \Bigg\langle  \delta_\alpha \left( \bk , \tau \right)  {\bar \delta}_\beta \left( \bp , \tau' \right) 
\sum_{m=1}^\infty \frac{1}{m!} \int d {\bf p'}_1 \dots  d {\bf p'}_m \left[ i \bq \cdot  \Delta{\bf D}_{\gamma\beta}({\bf p'}_1 ;\tau'',\tau')  \right] \dots \left[ i \bq \cdot  \Delta{\bf D}_{\gamma\delta}({\bf p'}_m ;\tau'',\tau')  \right] \nonumber\\
& &   {\bar \delta}_\gamma \left( \bq_1 - {\bf p'}_1 - \dots - {\bf p'}_m , \tau''  \right)  \Bigg\rangle \,,
\eeqra 
In the regime identified by \re{sql} and \re{displ}, and assuming that the displacement field is gaussian, the leading contribution to the expectation value above is given, at any order $m$,  by the one in which the two-point correlator between density fields at large momenta ($\langle\bar\delta_\beta \bar\delta_\gamma  \rangle$) factorizes, and the remaining correlator is further decomposed into a product of $2-$point correlators:  
\beqra
&&\!\!\!\!\!\! \!\!\!\!\!\!\ \!\!\!\!\!\!\!\!\!\!\!\! \!\!\!\!\!\!\!\!\!\!\!\!  B^{(\delta)}_{\alpha\beta\gamma}(\bk,\bp,\bq;\tau,\tau',\tau'') \delta_D \left( \bk + \bp + \bq \right) \nonumber\\
& & \!\!\!\!\!\! \!\!\!\!\!\!\ \!\!\!\!\!\!\!\!\!\!\!\! \!\!\!\!\!\!\!\!\!\!\!\!   \cong 
\sum_{n=1}^\infty \frac{\left(2n+1\right)\left(2n-1\right)!!}{\left(2n+1\right)!} \int d^3 \bp'_1 \dots  d^3 \bp'_{2n+1}  
\left\langle  \delta_\alpha \left( \bk , \tau \right) i \bq \cdot \Delta  {\bf D}_{\gamma \delta} \left( \bp'_{2w+1} ; \tau'', \tau' \right) \right\rangle \nonumber\\ 
& & \!\!\!\!\!\! \!\!\!\!\!\!\ \!\!\!\!\!\!\!\!\!\!\!\! \!\!\!\!\!\!\!\!\!\!\!\!  
 \left\langle {\bar \delta }_\beta \left( \bp, \tau' \right)   {\bar \delta}_\gamma \left( \bq - \bp'_1 - \dots - \bp'_{2w+1} , \tau'' \right) \right\rangle \prod_{\lambda=1}^n \left\langle 
 i \bq \cdot \Delta  {\bf D}_{\gamma\delta} \left( \bp'_{2\lambda-1} ; \tau'', \tau' \right) 
 i \bq \cdot \Delta  {\bf D}_{\gamma\delta} \left( \bp'_{2\lambda} ; \tau'', \tau' \right) \right\rangle\,, \nonumber\\ 
\label{B-partial}
\eeqra 
where we have taken into account that there are  $2n+1$ equivalent ways of contracting $ \delta_\alpha$ with one of the $\Delta  {\bf D}$, and $(2n-1)!!$ equivalent pairs of the remaining displacement vectors. 
If, in the limit of very large wavelength, $\bv_{\rm long}(\bk,\tau)$ can be expressed in terms of the linear field then,
\beq 
\!\!\!\!\!\!\!\! \!\!\!\!\!\!\!\!  \!\!\!\!\!\!\!\! 
 v^j_{\alpha,{\rm long}}(\bk,\tau) \simeq -i \frac{k^j}{k^2} \theta_{\alpha,L} (\bk,\tau) 
  \;\;\Rightarrow\;\;  i \bq \cdot \Delta  {\bf D}_{\alpha \beta} \left( \bk; \tau, \tau' \right)=   \frac{\bq \cdot \bk }{k^2}  \vartheta_{\alpha \beta} \left( \bk ; \tau, \tau' \right) \;,  
  \label{linearth}
\eeq
where we have defined 
\beqra
  \vartheta_{\alpha \beta} \left( \bk ; \tau, \tau' \right) &&\equiv   
   \int_{\tau_{in}}^\tau d \tau'' \theta_{\alpha,L} \left( \bk, \tau'' \right) -    \int_{\tau_{in}}^{\tau'} d \tau'' \theta_{\beta,L} \left( \bk, \tau'' \right)  \;, \nonumber\\
&& =  \frac{\theta_{\alpha,L} \left( \bk, \tau \right)}{{\cal H}(\tau) f(\tau)} - \frac{\theta_{\beta,L} \left( \bk, \tau' \right)}{{\cal H}(\tau') f(\tau')}\,,
\label{vartheta-def}
\eeqra
where  we have used the relation $d \tau =d D_+/({\cal H} f D_+)$. Moreover, we implemented the linear theory result  $\theta_L/({\cal H}f D_+)={\mathrm{const}}$, and we have considered the initial time so remote that the limits  $D_+ \left(\tau_{in}\right)/D_+ \left(\tau \right)\to 0$ and  $D_+ \left(\tau_{in}\right)/D_+ \left(\tau' \right)\to 0$ can be taken.

The approximation  in \re{linearth} becomes an equality in the limit in which the long wavelength mode converges to the linear theory result. Notice that, to have this convergence, the long-wavelength velocity should have a non-vanishing time dependence matching that of linear theory in the given cosmology. For instance, in Einstein-deSitter, it should grow linearly with $\tau$. 

Inserting this in (\ref{B-partial}), we obtain

\begin{eqnarray}
 \!\!\!\!\!\! \!\!\!\!\!\!\ \!\!\!\!\!\!\!\!\!\!\!\! \!\!\!\!\!\!\!\!\!\!\!\!  
\lim_{k\ll q} B^{(\delta)}_{\alpha\beta\gamma} \left( \bk, -\bq - \bk, \bq ; \tau, \tau', \tau'' \right) & \cong & - \frac{ \bq \cdot \bk }{ k^2 } 
{\rm e}^{-\frac{1}{2} \int  d^3 p' \left( \frac{ \bq \cdot \bp' }{ p'^2 }  \right)^2 \left\langle 
\vartheta_{\gamma\beta} \left(  \bp' ; \tau'', \tau' \right) 
\vartheta_{\gamma\beta} \left( - \bp' ; \tau'', \tau' \right) \right\rangle' } \nonumber\\ 
& & \quad  \times  
\left\langle  \delta_\alpha \left( \bk , \tau \right)  \vartheta_{\gamma\beta} \left( - \bk ; \tau'', \tau' \right) \right\rangle' 
{\bar P}_{\beta\gamma}^{(\delta)} \left( q ; \tau' , \tau'' \right) \,, 
\label{B-abc-par}
\end{eqnarray} 
where  the prime in $\langle \dots \rangle^\prime$ indicates that an overall momentum delta function has been factored out, and where 
\begin{equation} 
{\bar P}^{(\delta)}_{\beta\gamma} \left( q; \tau',\tau'' \right) \equiv \langle {\bar \delta}_\beta \left( - \bq , \tau' \right)  {\bar \delta}_\gamma \left(  \bq , \tau'' \right) \rangle' \;. 
\label{PS}
\end{equation} 

Performing the shift \re{filrep} in the two point correlator as we did for the bispectrum in \re{bispa}, and proceeding identically to \re{B-partial}, we can relate the PS of the barred and unbarred fields: 
\begin{equation} 
\!\!\!\!\!\!\!\! 
P^{(\delta)}_{\beta\gamma} \left( q; \tau',\tau'' \right) =  
{\bar P}^{(\delta)}_{\beta\gamma} \left( q; \tau',\tau'' \right)  {\rm e}^{-\frac{1}{2} \int  d^3 p' \left( \frac{ \bq \cdot \bp' }{ p'^2 }  \right)^2 \left\langle 
\vartheta_{\gamma\beta} \left(  \bp' ; \tau'', \tau' \right) 
\vartheta_{\gamma\beta} \left( - \bp' ; \tau'', \tau' \right) \right\rangle' } \;, 
\end{equation}
so that the exponential factor in \re{B-abc-par} is reabsorbed in the PS of the original (unbarred) fields, to give 
\begin{eqnarray}
 \!\!\!\!\!\! \!\!\!\!\!\!\ \!\!\!\!\!\!\!\!\!\!\!\! \!\!\!\!\!\!\!\!\!\!\!\!  
\lim_{k\ll q} B^{(\delta)}_{\alpha\beta\gamma} \left( \bk, -\bq - \bk, \bq ; \tau, \tau', \tau'' \right) & \cong & - \frac{ \bq \cdot \bk }{ k^2 } 
\left\langle  \delta_\alpha \left( \bk , \tau \right)  \vartheta_{\gamma\beta} \left( - \bk ; \tau'', \tau' \right) \right\rangle' 
 P_{\beta\gamma}^{(\delta)} \left( q ; \tau' , \tau'' \right) \,. \nonumber\\ 
\label{B-abc}
\end{eqnarray}

This relation is valid for an arbitrary number of species. In the reminder of this section we discuss it in the context of one or two species. In the single species case considered in  \cite{Peloso:2013zw}, 
\beq
  \vartheta \left( \bk; \tau, \tau' \right) =  \delta_L \left( \bk, \tau \right) \left[ \frac{D_+ \left( \tau' \right)}{D_+ \left( \tau \right)}-1 \right]  \,, \nonumber\\
\label{vartheta1}
\eeq
where we recall that the suffix $L$ indicates that the corresponding function is in the linear regime, and we have used the relation for the linear growing mode  $\theta_L\left( \bk, \tau \right) = - \delta_L \left( \bk, \tau \right) {\cal H}(\tau) f(\tau)$. 

Eq.~\re{B-abc} then reads 
\begin{eqnarray}
 & &  \!\!\!\!\!\! \!\!\!\!\!\!\ \!\!\!\!\!\!\!\!\!\!\!\! \!\!\!\!\!\!\!\!\!\!\!\! 
\lim_{k\ll q} B^{(\delta)} \left( \bk, -\bq - \bk, \bq ; \tau, \tau', \tau'' \right)  =  - \frac{  \bq \cdot \bk }{ k^2 } 
\frac{D_+ \left( \tau' \right) - D_+ \left( \tau'' \right)}{D_+ \left( \tau \right)}
       P_L^{(\delta)} \left( k; \tau,\tau \right) 
 P^{(\delta)} \left( q ;  \tau' , \tau'' \right) \,, \nonumber\\ 
\label{consistency1}
\end{eqnarray}  

We note that an equal sign, rather than  $\cong$, is present in eq. (\ref{consistency1}),  because the shift (\ref{filrep}) is a symmetry of the system (disregarding gradients of the long mode) in the single species case, and therefore the extra terms mentioned after eq.~\re{expa} are absent in this case. This consistency relation (\ref{consistency1}) was first obtained in  \cite{Peloso:2013zw,Kehagias:2013yd}.

As already mentioned in \cite{Peloso:2013zw,Kehagias:2013yd}, the consistency relation emerges as a consequence of invariance of the system with respect to  generalized Galilean invariance, {\it i.e.} invariance by time-dependent boosts  \cite{Peloso:2013zw}, which allows to reabsorb the long wavelength velocity mode by the field redefinition \re{filrep} . We stress that  the derivation above applies for fully non-linear fields. The only point where the linear theory is advocated is when imposing the matching \re{linearth} for the long velocity mode, which allows us to write the correlator of the long wavelength velocity field in terms of the linear power spectrum. Moreover, the gradient of the long velocity mode, that we have neglected since eq.~\re{conti}, would give contributions suppressed as $k^2/q^2$ with respect to \re{consistency1} in the squeezed ($k\ll q$) limit.

The derivation of eq.~\re{consistency1} discussed in this section clarifies its meaning: it comes from the fact that we are cross-correlating the same density field at different times, and, due to the non-trivial displacement between the two time-slices given by eq.~\re{vartheta1}, a non vanishing contribution to the bispectrum would emerge even if the fields  $ \delta(\bk,\tau)$ were gaussian. The effect vanishes at equal times, $\tau'=\tau''$ since the relative displacement \re{reldisp} vanishes in that limit if $\gamma=\beta$. 

In the multi-field case considered in this paper this is not necessarily the case,  since large scale modes affect differently two species which have a nontrivial velocity bias. To see this, let us consider the equal time limit $\tau''=\tau'$ of the  relation (\ref{B-abc}) in the case of two species. We also set $\alpha=1$ for definiteness.
In this case the relation (\ref{vartheta-def}) gives 
\begin{eqnarray}     
\vartheta_{\beta\gamma} \left( \bk; \tau' \tau' \right) & = & \frac{\theta_{\beta,L} \left( \bk , \tau' \right) - \theta_{\gamma,L} \left( \bk , \tau' \right) }{{\cal H} \left( \tau' \right) f \left( \tau' \right) }  \,.
\end{eqnarray} 

The density-velocity cross PS is then
\begin{equation}
 \!\!\!\!\!\! \!\!\!\!\!\!\ \!\!\!\!\!\!\!\!\!\!\!\! \!\!\!\!\!\!\!\!\!\!\!\! 
\left\langle { \delta}_1 \left( \bk, \tau \right) \frac{ \theta_{\beta,L} \left( - \bk,  \tau \right) }{{\cal H} \left( \tau \right) f \left( \tau \right) } 
\right\rangle' \equiv - P_{1\beta,L}^{(\delta v)} \left( k; \tau, \tau \right) = \left\{ \begin{array}{c} - P_{11,L}^{(\delta v)} \left( k; \tau, \tau \right) \;\;,\;\; \beta = 1 \\  - b_v \left( k,\tau \right) \, P_{11,L}^{(\delta v)} \left( k; \tau, \tau \right) \;\;,\;\; \beta = 2 \end{array} \right. \;, 
\label{P1beta}
\end{equation}
where $b_v \left( k,\tau \right)$ is the velocity bias between the two species  at the scale $k$.   The density-velocity  power spectrum is normalized in such a way that $P^{(\delta v)}  = P^{(\delta)}$ for a single species in the linear regime.   

The relation \re{B-abc} then gives, at $\tau'=\tau''$, 
\begin{eqnarray} 
  & &  \!\!\!\!\!\! \!\!\!\!\!\!\ \!\!\!\!\!\!\!\!\!\!\!\! \!\!\!\!\!\!\!\!\!\!\!\! 
\lim_{k\ll q} B^{(\delta)}_{1\beta\beta} \left( \bk, -\bq - \bk, \bq ; \tau, \tau', \tau' \right)  = 0 \,,  \nonumber\\   
  & &  \!\!\!\!\!\! \!\!\!\!\!\!\ \!\!\!\!\!\!\!\!\!\!\!\! \!\!\!\!\!\!\!\!\!\!\!\! 
\lim_{k\ll q} B^{(\delta)}_{112} \left( \bk, -\bq - \bk, \bq ; \tau, \tau', \tau' \right)  \cong   \left( 1- b_v \left( k ,\tau\right)  \right)  \frac{  \bq \cdot \bk }{ k^2 } \frac{D_+ \left( \tau' \right)     }{D_+ \left( \tau \right)}
  P_{11,L}^{(\delta v)} \left( k; \tau,\tau \right) 
 P_{12}^{(\delta)} \left( q ;  \tau' , \tau' \right) \,.   \nonumber\\ 
\label{consistency2}
\end{eqnarray}  

The result \re{consistency2} accounts for the contribution to the equal-time bispectrum from the kinematic effect (i) mentioned above. This contribution is exact in $q$ and in $1-b_v$.  The goal of the reminder of this work is to modify the relation  \re{consistency2} so to consistently include both the effects (i) and (ii) mentioned above.  Our final result is given in  eq. \re{nexwardbias}, which is valid for cross-correlators of density and/or velocity fields, and for arbitrary times.   One can imagine Taylor-expanding the  bispectrum in $1-b_v$. Each term in this Taylor expansion is a function of the hard and soft momenta, respectively $q$ and $k$. The second and third lines  of eq.  \re{nexwardbias} contain all the terms of this Taylor expansion due to (i), and - once specified to the density contrast and to equal time $\tau''=\tau'$ -  reproduce the result (\ref{consistency2}). The last  line at the right hand side (RHS) of eq.  \re{nexwardbias} contain the dominant term in $1-b_v$ due to (ii). It is remarkable, and highly nontrivial, that   also this term is exact in the nonlinear scale $q$, despite that it originates from the fact that the non-universal shift  \re{filrep} is not a symmetry of the system in this case. 

\section{Invariances of the tree level action}

\label{invariance-tree}
We will consider a system of two pressureless species, A, and B, described by the four-component field
\beq
\vp_a(\bk,\eta) =e^{-\eta} \left(
\begin{array}{c}
\delta_A(\bk,\eta)\\
-\frac{\theta_A(\bk,\eta)}{{\cal H} f}\\
 \delta_B(\bk,\eta)\\
-\frac{\theta_B(\bk,\eta)}{{\cal H} f}\\
\end{array}
\right)\,,
\label{multiplet}
\eeq
with $\eta\equiv \log \left[D_+(\tau)/D_+(\tau_{in})\right]$, 
 with $D_+$ and $\tau$ being the growth factor and the conformal time, respectively,  and where, as usual, we neglect the vorticities of the velocity fields.
We will consider equations of motion of the compact form \cite{Peloso:2013zw}
\begin{equation}
\!\!\!\!\!\!\!\!\!  \!\!\!  \left( \delta_{ab} \partial_\eta + \Omega_{ab} \right) \varphi_b \left( \bk , \eta \right) = {\rm e}^\eta 
\int d^3 \bp \, d^3 \bq \,  \gamma_{abc} \left( \bk ,\, - \bp ,\, - \bq \right) \varphi_b \left( \bp ,\, \eta \right) 
 \varphi_c \left( \bq ,\, \eta \right) \, , 
\label{compact}
\end{equation}
where
\begin{equation}
\ds {\bf \Omega} = \left( \begin{array}{cccc} 
1 & - 1 & 0 & 0\\
 \Omega_{21} &  \Omega_{22} &  \Omega_{23} & 0\\
0 & 0& 1& -1\\
 \Omega_{41} & 0 &  \Omega_{43} &  \Omega_{22}\\
\end{array} \right)\,,
\label{BIGOMEGA}
\end{equation}
and  where the only non vanishing components of the vertex function are
\begin{eqnarray}
&&
\gamma_{121} \left( \bk ,\, \bp ,\, \bq \right) = \frac{1}{2} \delta_D \left( \bk + \bp + \bq \right) \alpha \left( \bp , \bq \right)\,, \nonumber\\
&&
\gamma_{222} \left( \bk ,\, \bp ,\, \bq \right) =  \delta_D \left( \bk + \bp + \bq \right) \beta \left( \bp , \bq \right) \,\nonumber\\
&&
\gamma_{112} \left( \bk , \bp , \bq \right) = \gamma_{334} \left( \bk , \bp , \bq \right) = 
\gamma_{343} \left( \bk , \bq , \bp \right)= \gamma_{121} \left( \bk , \bq , \bp \right)   \;, \nonumber\\
&& \gamma_{444} \left( \bk ,\, \bp ,\, \bq \right)=\gamma_{222} \left( \bk ,\, \bp ,\, \bq \right)\,,
\label{vertex}
\end{eqnarray}
with $\delta_D$ being the Dirac $\delta-$function, and
\begin{equation}
\alpha \left( \bp, \bq \right) = \frac{\left( \bp+\bq \right) \cdot \bp}{p^2} \;\;,\;\;
\beta \left( \bp, \bq \right) = \frac{\left( \bp+\bq \right)^2 
\, \bp \cdot \bq  }{2 p^2 q^2} \, .
\label{al-be}
\end{equation}
The elements of the matrix ${\bf \Omega}$ at the first and third lines are fixed by the requirement that $A$ and $B$-type particles are separately conserved (continuity equation). In principle, the elements of the second and fourth lines (coming from the Euler equations) can be time and scale dependent and, moreover, the (44) component can be different from the (22) one . Such a situation can happen for instance, if the two species are coupled to a time-dependent scalar field with nonzero mass, see for instance \cite{ Am03, Saracco:2009df}. This would complicate the analysis without changing the results, and therefore we will consider a scale and time independent matrix {\bf $\Omega$}. Moreover, the structure of the vertex function, eq.~\re{vertex}, is fixed by the nonlinear terms in the continuity and Euler equations.

 We will stick to this general form of the equation for most of the paper, and discuss concrete physical systems governed by these equations in Sect.~\ref{examples}.

The equation of motion \re{compact} is invariant under the field transformation
\begin{equation}
\varphi_a \left( \bk , \eta \right) \rightarrow {\rm e}^{i \bk \cdot \bw( \eta)}
\varphi_a \left( \bk , \eta \right) + i  \, {\rm e}^{-\eta}  \; \partial_\eta \bw(\eta) \cdot \bk \;  \delta_D(\bk)\;d_a \, ,
\label{GT-phi}
\end{equation}
with $\bw(\eta)$ a uniform velocity field with arbitrary time dependence, and 
\beq
{\bf d}\equiv\left(\begin{array}{r}
0\\
1\\
0\\
1
\end{array}\right)\,.
\label{dd}
\eeq

In the special case in which 
\beq
\bw(\eta) = \bv_0\, T(\eta)\,,\qquad\qquad \mathrm{with}\qquad\quad T (\eta)\equiv \frac{1}{a(\eta)}\int_{0}^\eta d\eta^\prime \frac{a(\eta^\prime)}{f(\eta'){\cal H}(\eta^\prime)}\,,
\eeq
with $f=d \log D_+/d\log a$, the transformation above corresponds to a Galilean transformation ({\it i.e.} a boost by the constant velocity $\bv_0$ in physical coordinates), but, as we see, the invariance also holds for an acceleration transformation, that is, if we move to a non-inertial frame where apparent forces appear \cite{Peloso:2013zw,Kehagias:2013yd}. These velocity transformations are zero modes equally affecting all the particles, therefore they do not affect equal-time correlators between density and velocity fields. Another way to see it, is to realize that the zero modes apparent forces induced by an acceleration transformation would affect, through the Euler equation, only the zero modes of the velocity fields, leaving divergence $\nabla \cdot \bv$ and vorticity $\nabla \times \bv$ unaffected. The only nontrivial effect of this zero-mode transformation is the one through the vertex, and this is the same for acceleration and for simple GI.

\section{Ward identities}

\label{ward} 

As done in \cite{MP07b} we introduce the action
\begin{eqnarray}
&& S= S_{\rm free} + S_{\rm int} \nonumber\\
&&\;\;\;= \int d \eta d \etap  \,d^3 \bk \,\chi_a \left( - \bk , \eta \right) \, g_{ab}^{-1}(\eta,\etap)\, \varphi_b \left( \bk , \eta \right)\nonumber\\
&& \;\;\;- \int d \eta d^3 \bk  d^3 \bp \, d^3 \bq  {\rm e}^\eta  \,  \gamma_{abc} \left( - \bk ,\, - \bp ,\, - \bq \right) 
 \chi_a \left(  \bk , \eta \right)  \varphi_b \left( \bp ,\, \eta \right) 
 \varphi_c \left( \bq ,\, \eta \right),\nonumber\\
\label{action}
 \end{eqnarray}
where
\beq
 g_{ab}^{-1}(\eta,\etap) = \delta_D(\eta-\etap) \left( \delta_{ab} \partial_{\etap} + \Omega_{ab} \right)\,,
\label{gil}
 \eeq
is the inverse linear propagator, which, inverted imposing causal initial conditions, gives the linear propagator.

Extremizing the action in eq.~\re{action} with respect to $\chi_a$, gives the equation of motion \re{compact}. To complete the definition of the transformations \re{GT-phi}, 
we give the transformations for the fields $\chi_a$,
\beqra
&&\chi_a \left( \bk , \eta \right) \rightarrow {\rm e}^{i \bk \cdot \bw(\eta)  }
\chi_a \left( \bk , \eta \right) \, ,
\label{GT-chi}
\eeqra
so that, under (\ref{GT-phi}) and  (\ref{GT-chi}), 
\beqra
&& \!\!\!\!\!\!\!\!\!\! S \rightarrow S + \int d \eta d^3 \bk\,\bigg[ i  \chi_a \left( -  \bk , \eta \right) \left( \delta_{ab} \partial_\eta + \Omega_{ab} \right) d_b  \;{\rm e}^{-\eta}  \partial_\eta \bw(\eta) \cdot \bk\;\delta_D \left( \bk \right)\bigg] \, ,
\label{change-S}
\eeqra
where the  extra term gives a vanishing contribution to the equation of motion \re{compact} for $\bk\neq 0$. 

Let us consider a transformation by  an infinitesimal parameter $ \bw(\eta)$. Under this transformation, the fields $\varphi_a$ and $\chi_a$ transform as
the infinitesimal versions of \re{GT-phi} and \re{GT-chi}
\beqra
&& \delta\varphi_a \left( \bk , \eta \right) =  i \bk\cdot \bw(\eta)\, \varphi_a \left( \bk , \eta \right) 
+ i  \, {\rm e}^{-\eta}  \; \bk\cdot \partial_\eta \bw(\eta)  \;  d_a \,\delta_D(\bk)\,,\nonumber\\
&& \delta\chi_a \left( \bk , \eta \right) =  i \bk\cdot \bw(\eta) \;
\chi_a \left( \bk , \eta \right) \,.
\label{infsplitboost}
\eeqra

Let us consider the generating functional 
\begin{eqnarray}
Z \left[ J , K \right] & = & \int    {\cal D} \varphi \,   {\cal D} \chi \, {\rm exp } \Bigg\{ - \frac{1}{2} \int d^3 \bk\; \chi_a \left( - \bk , 0 \right) P_{ab}^0 \left( k \right) \chi_b \left( \bk , 0 \right) + i S \nonumber\\
& &  \!\!\!\!\!\!\!  \!\!\!\!\!\!\!  + i \int d \eta d^3 \bk \left[ J_a \left( - \bk , \eta \right) \varphi_a \left( \bk , \eta \right) +  K_a \left( - \bk , \eta \right) \chi_a \left( \bk , \eta \right) \right]  \Bigg\}    \nonumber\\
& \equiv &  \int    {\cal D} \varphi \,   {\cal D} \chi \, {\rm exp } \left\{ \dots \right\} \,, 
\label{Z}
\end{eqnarray}
and the change of variables (\ref{infsplitboost}) in this generating functional. The generating functional is invariant under any change of variable,  and using \re{change-S}, we get
\begin{eqnarray}
& &  \!\!\!\!\!\!\!\!  \!\!\!\!\!\!\!\! \!\!\!\!\!\!\!\!  \!\!\!\!\!\!\!\!  
  \delta Z = - 
\int   {\cal D} \varphi \,   {\cal D} \chi \, {\rm exp } \left\{ \dots \right\} \times\, 
  \nonumber\\
  &&   \!\!\!\!\!\!\!\!  \!\!\!\!\!\!\!\! \!\!\!\!\!\!\!\!  \!\!\!\!\!\!\!\!    \int d \eta d^3\bk\Bigg\{ \bk \cdot \bw(\eta) \; \bigg( J_a \left( - \bk , \eta \right) \varphi_a \left( \bk , \eta \right) +  K_a \left( - \bk , \eta \right) \chi_a \left( \bk , \eta \right) \bigg) \nonumber\\ 
  &&  \!\!\!\!\!\!\!\!  \!\!\!\!\!\!\!\! \!\!\!\!\!\!\!\!  \!\!\!\!\!\!\!\!  - \bk\cdot \bw(\eta)\; d_a \partial_\eta\Bigg[  {\rm e}^{-\eta}  \bigg( J_a \left( - \bk , \eta \right) + (-\delta_{ab}\partial_\eta+\Omega_{ab})\chi_b \left( - \bk ,\, \eta \right) \bigg)\Bigg]  \;\delta_D \left( \bk \right)\Bigg\}=0\,.
\label{dZ1}
\end{eqnarray}

The connected Green functions are obtained from functional derivatives of the functional  $W = -i \log Z$ with respect to the sources. Expressing the fields' expectation values, {\it i.e.}  the `classical' fields in presence of sources, as
\begin{equation}
{\tilde \vp}_a \left( \bk , \eta \right) \equiv  \frac{ \delta W }{ \delta J_a \left( - \bk , \eta \right) } \;\;,\;\;
{\tilde \chi}_a \left( \bk , \eta \right) \equiv  \frac{ \delta W }{ \delta K_a \left( - \bk , \eta \right) } \, ,
\label{field-sources}
\end{equation}
we get 
\beqra
&&\int d\eta\,d^3\bk\;\bw(\eta)\cdot \bk \Bigg\{J_a(-\bk,\eta)\frac{\delta W}{\delta J_a(-\bk,\eta)} + K_a(-\bk,\eta)\frac{\delta W}{\delta K_a(-\bk,\eta)} \nonumber\\
&&-d_a \partial_\eta\Bigg[ \mathrm{e}^{-\eta}\bigg(J_a(-\bk,\eta) +(-\delta_{ab}\partial_\eta+\Omega_{ab}) \frac{\delta W}{\delta K_b(\bk,\eta)}\bigg)\Bigg]\;\delta_D(\bk)\Bigg\}=0.
\label{WIconW}
\eeqra 
To obtain 1 particle irreducible (1PI) Green functions  one defines the effective action $\Gamma$, related to $W $ by   a Legendre transform:,
\begin{equation}
\!\!\!\!\!\!\!\!  \!\!\!\!\!\!\!\!  \!\!\!\!\!\!\!  \!\!\!\!\!\!\!\!  \Gamma \left[ {\tilde \varphi} , {\tilde \chi} \right] \equiv W \left[ J ,\, K \right] - \int d \eta d^3 \bk \left[ J_a \left( - \bk , \eta \right) {\tilde \varphi}_a \left( \eta ,\, \bf k \right) +  K_a \left( - \bk , \eta \right) {\tilde \chi}_a \left( \eta ,\, \bf k \right) \right] \,,
\end{equation}
where 
\begin{equation}
J_a \left( \bk , \eta \right) = - \frac{ \delta \Gamma }{ \delta {\tilde \varphi}_a \left( - \bk , \eta \right) } \;\;,\;\;
K_a \left( \bk , \eta \right) = - \frac{ \delta \Gamma }{ \delta {\tilde \chi}_a \left( - \bk , \eta \right) } \, ,
\label{JK-Gamma}
\end{equation}
and one then takes functional derivatives of the $\Gamma$ with respect to the classical fields. Using the above relations in the identity coming from the Galliean transformation, eq.~\re{WIconW}, we obtain the same identity as eq.~(45) of \cite{Peloso:2013zw},  namely
\begin{eqnarray}
& & 
  \int d \eta d^3 \bk \left( \bk \cdot \bw \right) \Bigg\{ T \left( \eta \right) \left[ 
    \frac{\delta \Gamma}{\delta \varphi_a \left( \bk  , \eta \right)} \varphi_a \left( \bk , \eta \right) +      \frac{\delta \Gamma}{\delta \chi_a \left( \bk  , \eta \right)} \chi_a \left( \bk , \eta \right) \right] \nonumber\\
& & + d_a\left[ \frac{\delta \Gamma}{\delta \varphi_a \left( \bk , \eta \right)} - \chi_b \left( - \bk ,\, \eta \right) \left( \delta_{ba} + \Omega_{ba} \right) \right] {\rm e}^{-\eta} \partial_\eta T \left( \eta \right) \delta_D \left( \bk \right) \Bigg\} = 0 \, ,
\label{dZ2}
\end{eqnarray}
where we have omitted the overtilde on the classical fields.
All the WI relating 1PI Green functions are obtained by taking functional derivatives of this expression, and then setting the fields to zero. These provide non-perturbative relations between terms of the effective action, as, for instance, eqs.~(47) and (51) of \cite{Peloso:2013zw}, to be satisfied in any viable approximation scheme.

\section{Consistency relations}

\label{consistency} 

In the previous section we have discussed exact relations coming from the transformation properties of the action, that is, of the equations of motion, under the uniform and {\it universal} ({\it i.e} species-independent) boost transformations \re{GT-phi}, \re{GT-chi}. In the derivation, the parameters of the transformations have been taken as arbitrary ones, as is usual done in the derivation of Ward identities, that is, we have not assumed that they correspond to physical long wavelength modes. The resulting Ward identities reflect the invariance of the equation of motions and provide local constraints on the structure of the mode-coupling interactions at a fully nonperturbative level. Notice that we did not have to assume any particular form for the tree-level mode-coupling term, but only that the action is invariant under generalized Galilean transformations, that is, eq.~\re{change-S}.

In this section, on the other hand, we will consider the effect of physical long wavelength velocity fields on the short scales. This will allow us to obtain CR for, {\it e.g.}, the bispectrum in the squeezed limit in the multi-species case, which can be straightforwardly generalized in CR for a generic $n-$point correlation function in the limit in which one of the incoming momentum is much smaller than all the others.

We split  this section in four parts. In the first part we perform an analogous transformation to the shift \re{filrep} in the partition function of the system. In the second part we derive the CR, in the last two parts we discuss it. 

\subsection{Effect of  the long scale velocity in the partition function} 

 The starting point is again the generating functional \re{Z}, where, following the discussion of Sect.~7 of  \cite{Peloso:2013zw}, we split the fields in long-- and short-- wavelength modes
\beq
\!\!\!\!\!\!\!\!\!\!\!\!  \!\!\!\!\!\! \vp_m(\bk,\eta)=\vp_m^{L}(\bk,\eta)+\vp_m^{S}(\bk,\eta)\,,\;\;\;\;\;\chi_m(\bk,\eta)=\chi_m^{L}(\bk,\eta)+\chi_m^{S}(\bk,\eta)\,,
\label{split2}
\eeq
where the splitting is defined by a proper window function that we do not need to specify here. 
Analogous splittings in long and short modes can be performed on the sources $J_m$ and $K_m$.

The  indices in $m,n,o,\cdots$ take values from $1$ to $4$  and, from now, on the indices $a,b,c$ are reserved for the fields that correspond to external lines in the final consistency relation that we formulate at the end of this Section.

Now the generating functional \re{Z} can be written as
\beqra
&&\!\!\!\!\!\!\!\!\!\!\!\!  \!\!\!\!\!\! \!\!\!\!\!\!\!\!\!\!\!\!  \!\!\!\!\!\!  Z[J,K] = \int {\cal D}\vp^{L} {\cal D}\chi^{L}  \, {\rm exp } \Bigg\{ - \frac{1}{2} \int d^3 \bk\, \chi_m^{L} \left( - \bk , 0 \right) P_{mn}^0 \left( k \right) \chi_n^{L} \left( \bk , 0 \right)+ i S^{L}[\vp^L_m,\chi^L_n]  \nonumber\\
& &  \!\!\!\!\!\!\!  \!\!\!\!\!\!\! \!\!\!\!\!\!\!\!\!\!\!\!  \!\!\!\!\!\!   + i \int d \eta d^3 \bk \left[ J_m^{L} \left( - \bk , \eta \right) \varphi_m^{L} \left( \bk , \eta \right) +  K_n^{L} \left( - \bk , \eta \right) \chi_n^{L} \left( \bk , \eta \right) \right]  \Bigg\}   \tilde Z[J^{S},K^{S};\vp^{L},\chi^{L}] \,,
\label{Zsplit}
\eeqra
where $S^{L}$ is obtained from the action \re{action} by replacing all the $\chi$ and $\vp$ fields with $\chi^L$ and $\vp^L$, respectively. The generating functional for the short-wavelength fields on the background of the long ones is given by
\beqra
&&\!\!\!\!\!\!\!\!\!\!\!\!  \!\!\!\!\!\! \!\!\!\!\!\!\!\!\!\!\!\!  \!\!\!\!\!\!  \tilde Z[J^{S},K^{S};\vp^{L},\chi^{L}] =  \int {\cal D}\vp^{S} {\cal D}\chi^{S}  \, {\rm exp }\Bigg\{ - \frac{1}{2} \int d^3 \bq\, \chi_m^{S} \left( - \bq , 0 \right) P_{mn}^0 \left( q \right) \chi_n^{S} \left( \bq , 0 \right) +  i S^{S}[\vp^S_m,\chi^S_n]\nonumber\\
&& -i \int d \eta d^3 \bk  d^3 \bp \, d^3 \bq  \,{\rm e}^\eta  \,  \bigg[ 
2 \, \gamma_{mno} \left( - \bq ,\, - \bp ,\, - \bk \right)  \chi_m^S \left(  \bq , \eta \right)  \varphi_n^S \left( \bp ,\, \eta \right)  \varphi_o^L \left( \bk ,\, \eta \right)\nonumber\\
 &&\;\;\;\;\;\;\;\;\;\;\;\;\;\;\;\;\;\;\;\;\;\;\;\;\;\;\;\;\;\;\;\;\;+  \gamma_{mno} \left( - \bk ,\, - \bp ,\, - \bq \right) \chi_m^L \left(  \bk , \eta \right)  \varphi_n^S \left( \bp ,\, \eta \right)  \varphi_o^S \left( \bq ,\, \eta \right) \bigg]\nonumber\\ 
& & + i \int d \eta d^3 \bq \left[ J_m^{S} \left( - \bq , \eta \right) \varphi_m^{S} \left( \bq , \eta \right) +  K_n^{S} \left( - \bq , \eta \right) \chi_n^{S} \left( \bq , \eta \right) \right]  \Bigg\}  \,,
\label{Zshort}
\eeqra
where $S^{S}$ is obtained from the action \re{action} by replacing all the $\chi$ and $\vp$ fields with $\chi^S$ and $\vp^S$, respectively.

From \re{Zshort} we see that the effect of the long modes on the short ones is mediated by the cubic terms at the second and third lines.  
 Moreover, from eq.~\re{vertex} we notice that in the $k\ll q$ regime the $\chi^L\vp^S\vp^S$ term at the third line of \re{Zshort} is suppressed at least as $k^2/q^2$ with respect to the $\chi^S\vp^S\vp^L$ at the second line, and therefore we will neglect it in the following.
 
Again exploiting the explicit expression of the vertex, we can approximate the second line of \re{Zshort} as
\beqra
&&-i \int d \eta d^3 \bk  d^3 \bp \, d^3 \bq  \,{\rm e}^\eta  \,  
2 \, \gamma_{mno} \left( - \bq ,\, - \bp ,\, - \bk \right)  \chi_m^S \left(  \bq , \eta \right)  \varphi_n^S \left( \bp ,\, \eta \right)  \varphi_o^L \left( \bk ,\, \eta \right)\nonumber \\
&&\simeq -i \int d\eta \int d^3\bp \,\Bigg\{ \chi_m^S \left( - \bp , \eta \right)  \varphi_m^S \left( \bp ,\, \eta \right) \,p^j \int d^3 \bk \,e^\eta \frac{k^j }{k^2} \, \frac{\vp_{2}^L(\bk,\eta) + \vp_{4}^L(\bk,\eta)}{2} +\nonumber\\
&&  \chi_m^S \left(  -\bp , \eta \right)h_{mn}  \varphi_n^S \left( \bp ,\, \eta \right) \,q^j \int d^3 \bk \,e^\eta \frac{k^j }{k^2} \, \frac{ \vp_{2}^L(\bk,\eta) - \vp_{4}^L(\bk,\eta) }{2}\Bigg\}+O((k/p)^0)\,
\eeqra
where, in order to compactify the notation we have introduced the matrix ${\bf h}$,
\beq
{\bf h}\equiv\left(\begin{array}{rrrr}
1&0&0&0 \\
0&1&0&0\\
0&0&-1&0\\
0&0&0&-1\\
\end{array}\right)\,.\qquad\qquad
\eeq

This expression rewrites
\begin{eqnarray}
- \int d \eta d^3 p \; p^j \chi_m^s \left( - \bp , \eta \right) \vp_m^s \left( \bp , \eta \right) \left[ D^j \left( \eta \right) + h_m^D \Delta D^j \left( \eta \right) \right] \,,
\label{iSint}
\end{eqnarray}
where   $h^D_m=(1,1,-1,-1)$, and where 
\beq
D^j(\eta)\equiv \frac{D^j_2(\eta)+D^j_4(\eta)}{2}\,,\qquad \Delta D^j(\eta)\equiv \frac{D^j_2(\eta)-D^j_4(\eta)}{2}\,,
\label{d1}
\eeq
with
\beq
D_n^j(\eta)\equiv i \int d^3 \bk e^\eta\frac{k^j}{k^2}\vp_n^L(\bk,\eta)\,\;\;\;\;\;\;\;\;\;\mathrm{for}\;\;\;n=2,4\,.
\label{d2}
\eeq

We define the shifted fields in momentum space as
\beqra
&&\bar\vp_m^S(\bp,\eta) \equiv \e^{i\bp\cdot{\bf D}_2(\eta)} \vp_m^S(\bp,\eta)\qquad(\mathrm{for}\;\;\;m=1,2)\,,\nonumber\\
&&\bar\vp_m^S(\bp,\eta) \equiv \e^{i\bp\cdot{\bf D}_4(\eta)} \vp_m^S(\bp,\eta)\qquad(\mathrm{for}\;\;\;m=3,4)\,,
\eeqra
or, in more compact notation, 
\beq
{\bar \vp}_m^S \left( \bp, \eta \right) \equiv {\rm e}^{i \bp\cdot({\bf D}(\eta)+h^D_m\Delta {\bf D}(\eta))}   \vp_m^S(\bp,\eta)\qquad(\mathrm{for}\;\;\;m=1,2,3,4)\,,
\label{shiftmom}
\eeq
 and   analogously for  the fields ${\bar \chi}_m$. In terms of these fields,  the generating functional in eq.~\re{Zshort} acquires the form 
\beqra
&&\!\!\!\!\!\!\!\!\!\!\!\!  \!\!\!\!\!\! \!\!\!\!\!\!\!\!\!\!\!\!  \!\!\!\!\!\!  \tilde Z[J^{S},K^{S};\vp^{L},\chi^{L}] =   \nonumber\\
&&\!\!\!\!\!\! \!\!\!\!\!\!\!\!\!\!\!\!  \!\!\!\!\!\!  \int {\cal D} \bar\vp^{S} {\cal D}\bar\chi^{S} {\rm exp }\Bigg\{ - \frac{1}{2} \int d^3 \bq\, \bar\chi_m^{S} \left( - \bq , 0 \right) P_{mn}^0 \left( k \right) \bar\chi_n^{S} \left( \bq , 0 \right)  \nonumber\\
&&+ i\left(  S^{S}[\bar\vp^S_m,\bar\chi^S_n]+  \Delta S^{S}[\bar\vp^S_m,\bar\chi^S_n;\Delta {\bf D}]\right)\nonumber\\
& &\!\!\!\!\!\! \!\!\!\!\!\!\!\!\!\!\!\!  \!\!\!\!\!\! + i \int d \eta\, d^3 \bq \Bigg[\sum_{m=1}^4 \e^{-i\bq\cdot({\bf D}(\eta)+h^D_m\Delta {\bf D}(\eta))} \left( J_m^{S} \left( - \bq , \eta \right) \bar \varphi_m^{S} \left( \bq , \eta \right)+K_m^{S} \left( - \bq , \eta \right) \bar\chi_m^{S} \left( \bq , \eta \right) \right)
  \Bigg] \Bigg\}\,,\nonumber\\
\label{Zshortnew}
\eeqra
where
\beqra
&& \!\!\!\!\!\!\!\!\!\!\!\!  \!\!\!\!\!\!  \!\!\!\!\!\!  \Delta S^{S}[\bar\vp^S_m,\bar\chi^S_n;\Delta {\bf D}]\equiv\int d\eta d^3\bq \bigg[ \Omega_{23} \bar\chi_2^S(-\bq,\eta)\bar\vp_3^S(\bq,\eta)\left(\e^{2 i \bq\cdot\Delta {\bf D}(\eta)}-1 \right)\nonumber\\
&&\;\;\;\;\;\;\;\;\;\;\;\;\;\;\;\;\;\;\;\;\;\;\;\;\;\;\;\;+\Omega_{41} \bar\chi_4^S(-\bq,\eta)\bar\vp_1^S(\bq,\eta)\left(\e^{-2 i \bq\cdot\Delta {\bf D}(\eta)}-1 \right)\bigg]\,.
\label{riscr}
\eeqra

In practice, the field transformation  removes the interaction term \re{iSint}, but introduces the phases in the last line of \re{Zshortnew} and the  term 
\re{riscr},  coming from the non-invariance of the action under \re{shiftmom} for $\Delta{\bf D}\neq 0$ if the equations of motions for the two fluids are coupled, that is, if  $\Omega_{23}\neq 0$ or $\Omega_{41}\neq 0$. The latter is precisely the effect we disregarded in Sect.~\ref{gaussian}.
Since the Jacobian of the transformation \re{shiftmom} is one, all the dependence of the generating functional for the short-wavelength fields on the long wavelength velocity fields $\vp^L_{2,4}$  is in the source terms and in $ \Delta S^{S}$, via the uniform displacement fields ${\bf D}$ and $\Delta{\bf D}$.

 We will take into account the effect of the non-invariance of the action perturbatively, that is, we will expand $\exp(i \Delta S^{S})$ in \re{riscr} as
\beqra
&& \!\!\!\!\!\!\!\!\!\!\!\!  \!\!\!\!\!\!  \!\!\!\!\!\! \exp\left(i \Delta S^{S}[\bar\vp^S_m,\bar\chi^S_n;\Delta{\bf D}]\right) =1-\int d\eta d^3\bq \;[{\bf h},{\bf \Omega}]_{mn} \bar\chi^S_m(-\bq,\eta)\bar\vp_n^S(\bq,\eta)\; \bq\cdot \Delta{\bf D}(\eta)\nonumber\\
&&\qquad\qquad\qquad\qquad +O(([{\bf h},{\bf \Omega}] \;\Delta{ D})^2)\,,
\label{iDS}
\eeqra
where  the only non-vanishing elements are $[{\bf h}, {\bf \Omega}]_{23}=2 \,\Omega_{23}$ and $[{\bf h}, {\bf \Omega}]_{41}=-2\, \Omega_{41}$, {\it i.e.}, are proportional to the terms coupling the two species.

\subsection{Derivation of the consistency relation}

The bispectrum in the squeezed limit is obtained by the triple derivative
\beqra
&& \!\!\!\!\!\!\!\!\!\!\!\!  \!\!\!\!\!\!  \!\!\!\!\!\! \!\!\!\!\!\!  B_{abc}^{L,S,S}(\bk,\bq,\bp;\eta,\eta',\eta'') \delta_D(\bk+\bq+\bp) = \frac{(-i)^3}{Z}\frac{\delta^3 Z[J,K]}{\delta J_a^L(\bk,\eta)\delta J_b^S(\bq,\eta')\delta J_c^S(\bp,\eta'')}\nonumber\\
&& \!\!\!\!\!\!\!\!\!\!\!\!  \!\!\!\!\!\!  \!\!\!\!\!\! \!\!\!\!\!\! = \langle \vp_a^L(\bk,\eta)\bar\vp_b^S(\bq,\eta')\bar\vp_c^S(\bp,\eta'') e^{-i \bq\cdot\left({\bf D}(\eta')-{\bf D}(\eta'')+h^D_b\Delta{\bf D}(\eta')-h^D_c\Delta{\bf D}(\eta'')\right)}\rangle\nonumber\\
&& \!\!\!\!\!\!\!\!\!\!\!\!  \!\!\!\!\!\!  \!\!\!\!\!\! \!\!\!\!\!\! - \int ds d^3\bp'\;[{\bf h},{\bf \Omega}]_{mn} \langle \vp_a^L(\bk,\eta)\bar\vp_b^S(\bq,\eta')\bar\vp_c^S(\bp,\eta'') e^{-i \bq\cdot\left({\bf D}(\eta')-{\bf D}(\eta'')+h^D_b\Delta{\bf D}(\eta')-h^D_c\Delta{\bf D}(\eta'')\right)}\ \nonumber\\
&& \qquad\qquad \bar\chi^S_m(-\bp',s)\bar\vp_n^S(\bp',s)\; \bp'\cdot \Delta{\bf D}(s)  \rangle + O(([{\bf h},{\bf \Omega}] \;\Delta{ D})^2)\,,
\label{bisq}
\eeqra
where,  here and in the following, the expectation values at rhs are evaluated setting  $\Delta S=0$. Moreover, in order to simplify the expression for the phases, we have used the momentum delta-function coming from the ensemble average, and the squeezed limit condition $k\ll q$. 

The first term at the RHS of \re{bisq} can be computed by performing a calculation completely analogous to the one of Sect.~\ref{gaussian}. Indeed, defining
\beq
{\bf E}_{bc}(\eta',\eta'')\equiv {\bf D}(\eta')-{\bf D}(\eta'')+h^D_b\Delta{\bf D}(\eta')-h^D_c\Delta{\bf D}(\eta'')\,,
\label{d5}
\eeq
the first term at the RHS of \re{bisq} can be expressed as
\beq
\!\!\!\!\!\! \!\!\!\!\!\! \sum_{n=0}^\infty \frac{(-i)^n}{n!} \langle \vp_a^L(\bk,\eta)\bar\vp_b^S(\bq,\eta')\bar\vp_c^S(\bp,\eta'') E_{bc}^{i_1}(\eta',\eta'') \cdots E_{bc}^{i_n}(\eta',\eta'') \rangle q^{i_1}\cdots q^{i_n}\,.
\eeq
At each order $n$,  and assuming ${\bf E}_{bc}(\eta',\eta'')$ to be gaussian, the leading term in the high momentum is of order $q/k \times(q^2<E_{bc}^2>)^m$ (where $m=(n-1)/2$, since only odd $n$ orders contribute), and is obtained by Wick contracting one of the $E_{bc}'s$ with the soft field $\vp_a^L(\bk,\eta)$, the two hard ones, $\bar\vp_b^S$ and $\bar\vp_c^S$ between themselves, and the remaining $E_{bc}'s$ among them, in pairs (only odd orders in $n$ contribute). Working out the same combinatorics as in Sect.~\ref{gaussian} we get
\beqra
&& \!\!\!\!\!\!\!\!  \!\!\!\!\!\!\!\! 
-i q^j\langle \vp_a^L(\bk,\eta) E_{bc}^{j}(\eta',\eta'') \rangle \langle \bar\vp_b^S(\bq,\eta')\bar\vp_c^S(\bp,\eta'')  \rangle \sum_{m=0}^\infty \frac{1}{m!}\left(-\langle E_{bc}^{i}E_{bc}^{k}  \rangle q^iq^k\right)^m\,\nonumber\\
&& \!\!\!\!\!\!\!\!  \!\!\!\!\!\!\!\! =-i q^j\langle \vp_a^L(\bk,\eta) E_{bc}^{j}(\eta',\eta'') \rangle \langle \bar\vp_b^S(\bq,\eta')\bar\vp_c^S(\bp,\eta'')  \rangle \exp\left(-\frac{\langle E_{bc}^{i}E_{bc}^{k}  \rangle}{2} q^iq^k\right)\,.
\eeqra
The second term at the RHS of \re{bisq} can be worked out similarly. The ensemble average gives
\beqra
&&\!\!\!\!\!\! \!\!\!\!\!\! \!\!\!\!\!\! \!\!\!\!\!\! \!\!\!\!\!\!  \langle \vp_a^L(\bk,\eta)\bar\vp_b^S(\bq,\eta')\bar\vp_c^S(\bp,\eta'') e^{-i \bq\cdot {\bf E}_{bc}(\eta',\eta'') }\bar\chi^S_m(-\bp',s)\bar\vp_n^S(\bp',s)\; \bp'\cdot \Delta{\bf D}(s)  \rangle \nonumber\\
&&\!\!\!\!\!\! \!\!\!\!\!\! \!\!\!\!\!\! \!\!\!\!\!\! \!\!\!\!\!\!=  p'^j \left[ \langle \vp_a^L(\bk,\eta) \Delta D^j(s)\rangle 
-  \langle \vp_a^L(\bk,\eta) \bq \cdot {\bf E}_{bc}(\eta',\eta'') \rangle \langle \bq \cdot {\bf E}_{bc}(\eta',\eta'') 
 \Delta D^j(s)\rangle \right]  \nonumber\\ 
& & \times   \exp\left(-\frac{\langle E_{bc}^{i}E_{bc}^{k} \rangle}{2} q^iq^k\right)  i 
{\cal T}_{mnbc}^{\chi\vp\vp\vp} \left( -\bp' , \bp' , \bq , \bp ; s,s,\eta',\eta'' \right) \, \delta_D \left( \bp + \bq \right)  \, ,  
\label{e3}
\eeqra
where we have defined
\begin{equation}
\!\!\!\!\!\!\!\! \!\!\!\!\!\!\!\! \!\!\!\!\!\!\!\! \!\!\!\!\!\!\!\! \!\!\!\!\!\!\!\!  
\langle \bar\chi^S_m(-\bp',s)\bar\vp_n^S(\bp',s) \bar\vp_b^S(\bq,\eta')\bar\vp_c^S(\bp,\eta'') \rangle \equiv i   
 {\bar {\cal T}}_{mnbc}^{\chi\vp\vp\vp} \left( -\bp' , \bp' , \bq , \bp ; s,s,\eta',\eta'' \right) \, \delta_D \left( \bp + \bq \right) \,. 
\end{equation} 

From now on, we can omit the overbar on the short wavelength fields, since the difference between correlators involving fields with and without the bar gives contributions which are higher order in the large scale displacement.

To proceed in the computation, we assume that the long modes can be treated in linear perturbation theory.
We introduce the {\it linear velocity bias} between the two species as 
\beq
b_v(k,\eta'')\equiv   \frac{P_{a4}(k;\eta',\eta'')}{P_{a2}(k;\eta',\eta'')}\,,
\label{d3}
\eeq
which is $a$-- and $\eta'$--independent in the linear regime.
The various ensemble averages appearing in the expressions above can be computed using \re{d1}, \re{d2}, and \re{d3}, which give
\beqra
&&\!\!\!\!\!\! \!\!\!\!\!\! \!\!\!\!\!\!  -iq^j\langle \vp_a^L(\bk,\eta)D^j(\eta') \rangle = -\e^{\eta'}\frac{1+b_v(k,\eta')}{2} \frac{\bq\cdot\bk}{k^2} \,P_{a2}^L(k;\eta,\eta')\,,\nonumber\\
&&\!\!\!\!\!\! \!\!\!\!\!\! \!\!\!\!\!\! -iq^j\langle \vp_a^L(\bk,\eta)\Delta D^j(\eta') \rangle = -\e^{\eta'}\frac{1-b_v(k,\eta')}{2} \frac{\bq\cdot\bk}{k^2} \,P_{a2}^L(k;\eta,\eta')\,,\nonumber\\
&&\!\!\!\!\!\! \!\!\!\!\!\! \!\!\!\!\!\! q^iq^j \langle D^i(\eta)D^j(\eta') \rangle = \e^{\eta+\eta'} \frac{q^2}{3} \int d^3 p \frac{P_{22}^L(p;\eta,\eta')}{p^2}\frac{(1+b_v(p,\eta))(1+b_v(p,\eta'))}{4}\,,\nonumber\\
&&\!\!\!\!\!\! \!\!\!\!\!\! \!\!\!\!\!\! q^iq^j \langle D^i(\eta)\Delta D^j(\eta') \rangle = \e^{\eta+\eta'} \frac{q^2}{3} \int d^3 p \frac{P_{22}^L(p;\eta,\eta')}{p^2}\frac{(1+b_v(p,\eta))(1-b_v(p,\eta'))}{4}\,,\nonumber\\
&&\!\!\!\!\!\! \!\!\!\!\!\! \!\!\!\!\!\! q^iq^j \langle \Delta D^i(\eta)\Delta D^j(\eta') \rangle = \e^{\eta+\eta'} \frac{q^2}{3} \int d^3 p \frac{P_{22}^L(p;\eta,\eta')}{p^2}\frac{(1-b_v(p,\eta))(1-b_v(p,\eta'))}{4}\,. \nonumber\\ 
\eeqra
Notice that, due to the filter function implicit in the splitting \re{split2} the above momentum integrals are limited to small momenta. Using the last three equations above and the definition \re{d5} we get
\beq
\exp\left(-\frac{\langle E_{bc}^{i}E_{bc}^{j}  \rangle}{2} q^iq^j\right) = \exp\left( -\frac{q^2 \sigma_{bc}^2 
\left( \eta', \eta'' \right) }{2} \right) \;,  
\eeq
with $\sigma_{bc}^2$ the long wavelength velocity dispersion

\beqra
& & \!\!\!\!\!\!\!\!  \!\!\!\!\!\!\!\!  \!\!\!\!\!\!\!\! 
\sigma_{bc}^2 \left( \eta', \eta'' \right)  \equiv   \frac{1}{3}\int d s ds' d^3\bp' \frac{P_{22}^L(p';s,s')}{p'^2}\,
{\cal F}_{bc} \left[ p' ; s, \eta', \eta'' \right] {\cal F}_{bc} \left[ p' ; s', \eta', \eta'' \right]  \;,  \nonumber\\ 
& &  \!\!\!\!\!\!\!\!  \!\!\!\!\!\!\!\!  \!\!\!\!\!\!\!\! 
{\cal F}_{bc} \left[ p'; s, \eta', \eta''\right]  \equiv  
 \e^{\eta'} \left(\frac{1+b_v(p',\eta')}{2} +\frac{1-b_v(p',\eta')}{2} h^D_b\right) \delta_D \left( s - \eta' \right) \nonumber\\ 
 & & \quad\quad\quad\quad - \left( \eta'\to \eta'',\;b\to c \right) \;.     
\eeqra

We also get
\beqra
&&\!\!\!\!\!\! \!\!\!\!\!\! \!\!\!\!\!\! 
  p'^j \left[ \langle \vp_a^L(\bk,\eta) \Delta D^j(s)\rangle 
-  \langle \vp_a^L(\bk,\eta) \bq \cdot {\bf E}_{bc}(\eta',\eta'') \rangle \langle \bq \cdot {\bf E}_{bc}(\eta',\eta'') 
 \Delta D^j(s)\rangle \right]  \nonumber\\ 
&&\!\!\!\!\!\! \!\!\!\!\!\!   =  - i {\rm e}^s  \left\{ \frac{1-b_v \left( k,s \right)}{2} \frac{\bp' \cdot \bk}{k^2}P^L_{a2} \left( k;\eta,s \right) - \frac{\bq \cdot \bk}{k^2} \frac{\bq \cdot \bp'}{3} 
{\cal Q}_{a} \left[ k; s, \eta,  \eta', \eta'' \right] \right\} \,,  
\label{longcont-term2}
\eeqra
where we have defined the quantity (disregarding in a consistent way terms of higher order in $1-b_v$)
\beqra
&&\!\!\!\!\!\! \!\!\!\!\!\! \!\!\!\!\!\!  \!\!\!\!\!\! 
{\cal Q}_{a} \left[ k;s, \eta,\eta', \eta'' \right]   \equiv   
\left(  {\rm e}^{\eta'} P^L_{a2}(k;\eta,\eta') - \eta' \leftrightarrow \eta''  \right) \nonumber\\ 
& & \qquad\qquad \qquad \times 
\int \frac{d^3 p''}{p''^2} \frac{1-b_v \left( p'',s \right)}{2} \; 
\left(  {\rm e}^{\eta'} P^L_{22}(p'';s,\eta') - \eta' \leftrightarrow \eta''  \right)  \;.   
\eeqra
Putting all together, we find 
\beqra
&&\!\!\!\!\!\!\!\!\!\!\!\!\!\! \!\!\!\!\!\!\!\!\!\!\!\!\!\!\!\! \lim_{k\ll q}  B_{abc}^{L,S,S}(k,q,|\bq+\bk|;\eta,\eta',\eta'') =
  \exp\left( -\frac{q^2 \sigma_{bc}^2 \left( \eta' , \eta'' \right)}{2}  \right) \nonumber\\ 
&&\!\!\!\!\!\!\!\!\!\!\!\!\!\! \!\!\!\!   \Bigg\{ - \frac{\bk \cdot \bq}{k^2}
\Big[   \frac{1+b_v \left( k ,\eta\right)}{2} \left( {\rm e}^{\eta'} P^L_{a2}(k;\eta,\eta')- 
\eta' \leftrightarrow \eta'' \right) {\bar P}_{bc} \left( q; \eta',\eta'' \right) \nonumber\\
& & \quad + \frac{1-b_v \left( k ,\eta\right)}{2} \left(  e^{\eta'}h_{bl} {\bar P}_{lc} \left(q;\eta',\eta'' \right) P^L_{a2}(k;\eta,\eta') -
\left( \eta' \leftrightarrow \eta'' ; b \leftrightarrow c \right)
  \right) \Big]\nonumber\\
&&\!\!\!\!\!\!\!\!\!\!\!\!\!\! \!\!\!\!\!\! + \int d^3 \bp'  \int d s {\rm e}^s \left[ {\bf h},{\bf \Omega} \right]_{mn}  
 \left[ \frac{1-b_v \left( k,\eta \right)}{2} \frac{\bp' \cdot \bk}{k^2}P^L_{a2}(k;\eta,s) - \frac{ \bq \cdot \bk }{ k^2 } \frac{\bq \cdot \bp' }{3} {\cal Q}_{a} \left[ k; s, \eta,  \eta', \eta'' \right] \right] \nonumber\\ 
& & \quad\quad  \quad\quad  \quad\quad 
   {\bar  {\cal T}}^{\chi\vp\vp\vp}_{mnbc} \left( \bp',-\bp',\bq,-\bq;s,s,\eta',\eta'' \right)\,\Bigg\} +O(k^0)\,.
\label{nexwardbias-phase}
\eeqra
where ${\bar P}$ is the power spectrum of the shifted fields ${\bar \vp}^S$. Using the relation \re{shiftmom}, it is immediate to see that 
\begin{equation}
\!\!\!\!\!\!\!\! \!\!\!\!\!\!\!\! \!\!\!\!\!\!\!\! \!\!\!\!\!\!\!\! \!\!\!\!\!\!\!\! 
P_{bc} \left( q ; \eta', \eta'' \right) = \langle \bar\vp_b^s \left(  \bq, \tau' \right)  \bar\vp_c^s \left( - \bq, \tau'' \right) 
{\rm e}^{-i \bq \cdot {\bf E}_{bc}(\eta',\eta'')} \rangle' = 
  \exp\left( -\frac{q^2 \sigma_{bc}^2 \left( \eta' , \eta'' \right)}{2}  \right) {\bar P}_{bc} \left( q ; \eta', \eta'' \right) \,, 
\end{equation}
and analogously for the four point function 
\begin{equation}
\!\!\!\!\!\!\!\! \!\!\!\!\!\!\!\! \!\!\!\!\!\!\!\! \!\!\!\!\!\!\!\! \!\!\!\!\!\!\!\! 
 {\cal T}^{\chi\vp\vp\vp}_{mnbc} \left( \bp',-\bp',\bq,-\bq;s,s,\eta',\eta'' \right)\, 
 =  \exp\left( -\frac{q^2 \sigma_{bc}^2 \left( \eta' , \eta'' \right)}{2}  \right) 
  {\bar {\cal T}}^{\chi\vp\vp\vp}_{mnbc} \left( \bp',-\bp',\bq,-\bq;s,s,\eta',\eta'' \right) \,. 
\end{equation}
Namely, on the RHS of \re{nexwardbias-phase} the exponential prefactor  is reabsorbed and we  finally get  the consistency relation for the squeezed bispectrum (eq.~\re{bisq}) in presence of large scale velocity bias, in terms of the original (unshifted) fields:  
\beqra
&&\!\!\!\!\!\!\!\!\!\!\!\!\!\! \!\!\!\!\!\!\!\!\!\!\!\!\!\!\!\! \!\!\!\!\!\!\! \lim_{k\ll q}  B_{abc}^{L,S,S}(k,q,|\bq+\bk|;\eta,\eta',\eta'') =
 \nonumber\\ 
&&\!\!\!\!\!\!\!\!\!\!\!\!\!\! \!\!\!\!    - \frac{\bk \cdot \bq}{k^2}  \Bigg\{
   \frac{1+b_v \left( k ,\eta\right)}{2} \left( {\rm e}^{\eta'} P^L_{a2}(k;\eta,\eta')- 
\eta' \leftrightarrow \eta'' \right)  P_{bc} \left( q; \eta',\eta'' \right) \nonumber\\
& & \quad + \frac{1-b_v \left( k ,\eta\right)}{2} \left(  e^{\eta'}h_{bl}  P_{lc} \left(q;\eta',\eta'' \right) P^L_{a2}(k;\eta,\eta') -
\left( \eta' \leftrightarrow \eta'' ; b \leftrightarrow c \right)
  \right) \nonumber\\
&&\!\!\!\!\!\!\!\!\!\! \!\!\!\!\!\!\!\!\!\!\!\!\!\!\!\! \!\!\!\!\!\!\!\!\!- \int d s {\rm e}^s \left[ {\bf h},{\bf \Omega} \right]_{mn}  
 \left[ \frac{1-b_v \left( k,\eta \right)}{2} P^L_{a2}(k;\eta,s) -  \frac{q^2   }{3} {\cal Q}_a \left[ k; s, \eta,  \eta', \eta'' \right] \right] {\cal I}_{mnbc}(q;s,\eta',\eta'')\Bigg\}\,,\nonumber\\
 &&\qquad\qquad\qquad\qquad\qquad\qquad\qquad\qquad\qquad\qquad +O(k^0)\,,
\label{nexwardbias}
\eeqra
where we exploited rotational invariance to rewrite the original $\bp'$ integral as
\begin{eqnarray}
\!\!\!\!\!\!\!\!\!\!\!\!\!\! \!\!\!\!\!\!\!\!\!\!\!\!\!\!\!\!  \!\!\!\!\!\!\!\!\!\!\!\!\!\!\!\! 
\int d^3 p' \bp'^i  {\cal T}^{\chi\vp\vp\vp}_{mnbc} \left( \bp',-\bp',\bq,-\bq;s,s,\eta',\eta'' \right) & = &  \bq^i 
\int d^3 p' \, \frac{  \bp' \cdot \bq }{q^2}  \, {\cal T}^{\chi\vp\vp\vp}_{mnbc} \left( \bp',-\bp',\bq,-\bq;s,s,\eta',\eta'' \right)  
\nonumber\\ 
& \equiv &  \bq^i \, {\cal I}_{mnbc}(q;s,\eta',\eta'') \;.
 \label{calf}
\end{eqnarray}

Notice that in the above expressions we never made the assumption that the linear mode $k$ is on the growing mode. If it were the case, the above expressions would further simplify by using the expression $P^L_{ab}(k;\eta,\eta') = u_a u_b P^0(k)$. Our results hold in the more general case in which isocurvature modes are present and, for the  baryon-DM fluid, they can accommodate the relative velocity between baryons and DM discussed  in \cite{Tseliakhovich:2010bj}.

\subsection{Discussion of the consistency relation at generic times}

Equation \re{nexwardbias}  is the result we were looking for. It gives the leading terms in the $k\ll q$ limit for the exact bispectrum in the multi field case, allowing for velocity bias between different species.   As discussed in Sect. \ref{gaussian} a nonvanishing bispectrum arises because of  (i)   the different large scale motion of the two species and (ii)  the coupling between the different species. The second and the third line of eq.  \re{nexwardbias} account for the effect (i), while the last line accounts for the effect (ii). To obtain the last  line,  we expanded linearly in $1-b_v(k)$, while  no such expansion has been performed in the terms that account for (i). It is immediate to verify that these terms reproduce exactly eqs. \re{consistency1} and \re{consistency2} given in Sect.  \ref{gaussian}, where only the effect (i) was accounted for. 

In the case of no bias, $b_v = 1$, eq. \re{nexwardbias} reduces to 
\beqra
&&\!\!\!\!\!\!\!\! \!\!\!\!\!\!\!\! \!\!\!\!\!\!\!\! \!\!\!\!\!\!\!\!  \!\!\!
 \lim_{k\ll q}  B_{abc}^{L,S,S}(k,q,|\bq+\bk|;\eta,\eta',\eta'') = 
 \;\frac{\bk\cdot\bq}{k^2}   \, P_{bc}(q;\eta',\eta'')\, u_2u_a  \,P^0(k)  \bigg(\mathrm{e}^{\eta'}-\mathrm{e}^{\eta''}  \bigg)\,, \quad(b_v = 1) \,, \nonumber\\ 
\label{wardnobias}
\eeqra
which (due to $b_v=1$) is a trivial generalization of the single species result, and indeed it reproduces   eq. \re{consistency1} in the single species case. 

As we discuss in Sect.~\ref{EPDM}, EP violation generally induces a velocity bias at arbitrarily large scales. At scales smaller than the horizon at decoupling, {\it i.e.} for $k> k_{dec}$ a relative velocity between baryons and DM as a consequence of the different post-decoupling initial conditions has been discussed in \cite{Tseliakhovich:2010bj}, where its effect on the formation of the first structures at small scales has been investigated. In Sect.~\ref{HaloDM} we discuss the velocity bias  established between the DM fluid  and the (proto) DM halos.  For this system, the  velocity bias is induced by a statistical selection effect and the identity \re{wardnobias} gets further  contributions, non vanishing at equal times and still proportional to $\bq \cdot \bk$. While the angular dependence of these terms is the same as for those induced by EP violation, the scaling with $k$ is different, being subdominant as $k\to 0$, which gives a potential handle to observationally separate the different origins of deviations from eq.~\re{wardnobias}.

Ref.~\cite{Creminelli:2013mca} derived  CR for the density contrast correlators starting from a relativistic description of the dark matter system. Their relations are obtained from diffeomorfism invariance, and reduce to eq.  \re{wardnobias}  \cite{Peloso:2013zw,Kehagias:2013yd}   in the non-relativistic regime. While ref.~\cite{Creminelli:2013mca} emphasized the role of the EP in obtaining these relations,  we think that the derivations of Sect.~\ref{gaussian} and of the present Section elucidate their origin as a purely kinematical effect, due to the mismatch between the two uniform shifts, eq.~ \re{filrep} or 
eq.~\re{shiftmom}, needed to erase the effect of a long wavelength velocity mode at the two different times corresponding to the short-distance points in the bispectrum. The non-equivalence between the CR \re{wardnobias} and the EP can be seen only in the multi species case. Indeed, has we have reviewed above, while for $k<k_{dec}$ a velocity bias can be induced only by EP violation, at smaller scales it can be produced also by other mechanisms, and if there is enough separation of scales, the modified CR derived in this paper hold also in these cases in which there is no EP violation.  It is nontrivial that fully nonlinear and possibly testable CR can be obtained also in this second case, extending the results of   \cite{Peloso:2013zw,Kehagias:2013yd,Creminelli:2013mca}.  Moreover, even if EP is violated, but GI is still an invariance of the equations of motion, the constraints on the effective action imposed by the identities derived from eq.~\re{dZ2} still hold.

Unfortunately, the test of the CR \re{nexwardbias} can 
unlikely be performed with observations only, as the quantity ${\cal T}$ can hardly be measured in the nonlinear regime by an observer who has access only to the light cone. Indeed, even in the $\eta'=\eta''$ case, the ${\cal T}$ term cannot be expressed using only equal time correlators. We can gain some insight on this term by expanding it in a  disconnected plus a connected (trispectrum) part 
\beqra
&&\!\!\!\!\!\! \!\!\!\!\!\! \!\!\!\!\!\! \!\!\!\!\!\! \!\!\!\!\!\!   \!\!\!\!\!\!   \!\!\!\!\!\!  
  {\cal T}^{\chi\vp\vp\vp}_{mnbc}(\bp',-\bp',\bq,-\bq;s,s,\eta',\eta'') \delta_D \left( \bp + \bq \right)  =  
\delta_D \left( \bp + \bq \right) \Bigg\{  G_{bm} \left( q ; \eta' , s \right) P_{nc} \left( q; s, \eta'' \right) \delta_D \left( \bp' + \bq \right) \nonumber\\ 
&&\!\!\!\!\!\! \!\!\!\!\!\! \!\!\!\!\!\! \!\!\!\!\!\! \!\!\!\!\!\!        +   G_{cm} \left( q ; \eta'' , s \right) P_{nb} \left( q; s, \eta' \right) \delta_D \left( \bp' - \bq \right)  +  T^{\chi\vp\vp\vp}_{mnbc}(\bp',-\bp',\bq,-\bq;s,s,\eta',\eta'') \Bigg\}\,, 
\eeqra
where we have defined the nonlinear two-point correlator, {\it i.e.} the PS, 
\beq
 \!\!\!\!\! \!\!\!\!\!\!\! \!\!\!\!\!\!\! \!\!\!\!\!\!\! \!\!\!\!\!\!\!\left.\frac{\delta^2 W}{\delta J_m(\bk,\eta)\delta J_n(\bk',\eta') } \right|_{J_m=K_n=0} = i \delta_D(\bk+\bk') P_{mn}(k;\eta,\etap)\,,
 \label{PSdef}
\eeq
the nonlinear propagator,
\beq
 \!\!\!\!\! \!\!\!\!\!\!\! \!\!\!\!\!\!\! \!\!\!\!\!\!\! \!\!\!\!\!\!\!\left.\frac{\delta^2 W}{\delta J_m(\bk,\eta)\delta K_n(\bk',\eta') } \right|_{J_m=K_n=0} = - \delta_D(\bk+\bk') G_{mn}(k;\eta,\etap)\,,
\label{propdef}
\eeq
and the connected four-point function
\beqra
&&\!\!\!\!\! \!\!\!\!\!\!\! \!\!\!\!\!\!\! \!\!\!\!\!\!\! \!\!\!\!\!\!\!\left.\frac{\delta^4 W}{\delta K_l(\bk,\eta) \delta J_m(\bk',\eta')\delta J_n(\bk'',\eta'')\delta J_o(\bk''',\eta''') } \right|_{J_m=K_n=0} =\nonumber \\
&& \qquad\qquad \delta_D(\bk+\bk'+\bk''+\bk''')\,T_{lmno}^{\chi\vp\vp\vp}(\bk,\bk',\bk'',\bk''';\eta,\etap,\eta'',\eta''')\,.
\label{trisp}
\eeqra
In this way, the expression \re{nexwardbias} rewrites 
\beqra
&&\!\!\!\!\!\!\!\!\!\!\!\!\!\! \!\!\!\!\!\!\!\!\!\!\!\!\!\! \lim_{k\ll q}  B_{abc}^{L,S,S}(k,q,|\bq+\bk|;\eta,\eta',\eta'') = - \frac{\bk \cdot \bq}{k^2}
 u_2 u_a P^0\left( k \right) \nonumber\\ 
&&\!\!\!\!\!\!\!\!\!\!\!\!\!\!    \Bigg\{ 
   \frac{1+b_v \left( k,\eta \right)}{2} \left( {\rm e}^{\eta'} - {\rm e}^{\eta''} \right) P_{bc} \left( q; \eta',\eta'' \right) \nonumber\\
& & \qquad + \frac{1-b_v \left( k,\eta \right)}{2} \left(  h_{bl} P_{lc} \left(q;\eta',\eta'' \right) e^{\eta'} - P_{bl} \left(q;\eta',\eta'' \right) h_{lc} e^{\eta''} \right) \nonumber\\
&&\!\!\!\!\!\!\!\!\!\!\!\!\!\! 
+   \int d s {\rm e}^s \left[ {\bf h},{\bf \Omega} \right]_{mn} 
   \left[ \frac{1-b_v \left( k,\eta \right)}{2} - \frac{q^2}{3}  {\cal Q} \left[    s, \eta', \eta'' \right] \right] \nonumber\\ 
& & \quad\quad \qquad    \left[ G_{bm} \left(q;\eta',s \right) P_{nc} \left( q;s,\eta'' \right) - G_{cm} \left( q;\eta'',s \right) P_{nb} \left( q;s,\eta' \right) \right] \nonumber\\ 
&&\!\!\!\!\!\!\!\!\!\!\!\!\!\!  -   \int d s {\rm e}^s \left[ {\bf h},{\bf \Omega} \right]_{mn} 
 \left[ \frac{1-b_v \left( k ,\eta\right)}{2}  -  \frac{q^2   }{3}  {\cal Q} \left[   s, \eta', \eta'' \right] \right] I_{mnbc}(q;s,\eta',\eta'') \,\Bigg\}\nonumber\\
 &&\qquad\qquad\qquad \qquad\qquad\qquad \qquad\qquad\qquad +O(k^0)\,,
\label{nexwardbias2}
\eeqra
where we have defined
\begin{equation}
\!\!\!\!\!\!\!\! \!\!\!\!\!\!\!\! \!\!\!\!\!\!\!\! 
{\cal Q} \left[ s, \eta', \eta'' \right]   \equiv   
\frac{{\cal Q}_{a} \left[ k;s, \eta,\eta', \eta'' \right] }{u_2 u_a P^0(k)} = 
\left(  {\rm e}^{\eta'} -  {\rm e}^{\eta''} \right)^2 u_2^2 
\int \frac{d^3 p''}{p''^2} \frac{1-b_v \left( p'',s \right)}{2} \; 
 P^0(p'')    \;,    
\end{equation}
and, analogously to \re{calf},
\beq
I_{mnbc}(q;s,\eta',\eta'')\equiv  \int d^3 \bp'   \frac{ \bp' \cdot \bq }{q^2} 
  T^{\chi\vp\vp\vp}_{mnbc} \left( \bp',-\bp',\bq,-\bq;s,s,\eta',\eta'' \right),
\eeq
where we recall that all the integrals are limited to small (linear)  momenta.
We see that the RHS of \re{nexwardbias2} contain propagators, PS, and the trispectrum  evaluated at different time arguments. As for the propagator, all these contributions involving unequal times cannot be directly measured, and should be computed in some (semi analytical) approximation scheme or in N-body simulations.   For example, we immediately see   that the 
term with the disconnected part of  ${\cal T}$  (the fourth and fifth line in  \re{nexwardbias2}) dominates over the term with the connected part (the last line in  \re{nexwardbias2})  in the linear and quasi-linear regime, 
as the connected part vanishes at tree level. In \ref{treetest} we check explicitly  the relation \re{nexwardbias2} at tree level. 
 
 Analogously to      the propagator
\begin{equation}
G_{ab} \left( k; \eta, \eta' \right)  = \left\langle \frac{ \delta \vp_a \left( \bk, \eta \right) }
{ \delta \vp_b \left( \bk, \eta' \right) } \right\rangle'  \,, 
\label{G-def2}
\end{equation}
(we remind that the prime simply indicates that a $\delta_D$ function has been factored out the expectation value)  the four-point function ${\cal T}^{\chi\vp\vp\vp}$ can be obtained from the response of the three point function $\langle \vp^3 \rangle$  to a change of the value of the field at the time carried by $\chi$    
\begin{equation}
\!\!\!\!\!\!\!\! \!\!\!\!\!\!\!\! 
 {\cal T}_{mnbc}^{\chi\vp\vp\vp} \left(  \bp', - \bp', \bq, - \bq ; s, s, \eta', \eta'' \right)     =      \left\langle \frac{ \delta \left(  \vp_n \left( - \bp', s \right)   \vp_b \left( \bq, \eta' \right)   \vp_c \left( - \bq, \eta'' \right) \right)  }{ \delta \vp_m \left( - \bp', s \right) } \right\rangle'   \,, 
\label{calT-def2}
\end{equation} 
and this formulation, done in terms of $\vp$ only, may be then used  to extract the four point function from an N-body simulation, where only the density ($\propto \vp_{1,3}$) and velocity ($\propto \vp_{2,4}$) fields appear.

\subsection{Discussion of the consistency relation at equal time $\eta' = \eta''$}

It is worth noting that the  equal time squeezed bispectrum has not the  ${\rm O } \left( \frac{\bq\cdot\bk}{k^2} \right)$ behavior for  $b=c$, namely
\begin{equation}
 \lim_{k\ll q}  B_{abb}^{\vp\vp\vp}(k,q,|\bq+\bk|;\eta,\eta',\eta') = {\rm O } \left( k^0 \right) \;. 
 \label{eq45}
 \end{equation}
To see this, we first show that, in total generality, 
\begin{equation}
\lim_{k\ll q} B_{abc} \left( \bk, \bq , - \bk - \bq ; \eta, \eta', \eta'' \right) = 
\lim_{k\ll q} B_{abc} \left(  \bk, - \bk + \bq ,  - \bq ; \eta, \eta', \eta'' \right) \;\;. 
\label{identity-B}
\end{equation}
To prove \re{identity-B}, we first note that the time and species labels  at LHS and RHS coincide. The same is true for the first side $\bk$. Finally, it is immediate to see that, in the squeezed limit,   the angle between any of the sides of the triangle at LHS is equal to the corresponding angle  at RHS. Therefore the two triangles coincide in the squeezed limit, and the identity  \re{identity-B} is proven. Evaluating this identity for $b=c$ and $\eta'=\eta''$, we can further interchange the second and the third side at RHS, and we then see that  the equal-time and equal-species squeezed bispectrum \re{eq45} does not change under $\bq \to - \bq$ (while $\bk$ is kept fixed).  Therefore, no   ${\rm O } \left( \frac{\bq\cdot\bk}{k^2} \right)$ contribution is present, and hence eq. \re{eq45} is valid.  Incidentally, we can also see by direct inspection that the our expression  \re{nexwardbias} for the bispectrum satisfy both  \re{eq45} and  \re{identity-B}. 

The $b \neq c$ case is more interesting.  While the second line  of \re{nexwardbias}  (the only term present at  $b_v=1$)  vanishes at $\eta' = \eta''$, this is not necessarily the case for the remaining  terms.  Hence, the study of the  $\eta' = \eta''$ squeezed bispectrum can provide information on the presence of a nontrivial  velocity bias between different species in the theory. From eq.~ \re{nexwardbias}, the equal time squeezed bispectrum reads
\beqra
&&\!\!\!\!\!\!\!\!\!\!\!\!\!\! \!\!\!\!\!\!\!\!\!\!\!\!\!\!  \!\!\!\!\!\!\!\!\!\!\!\!\!\! 
\lim_{k\ll q}  B_{abc}^{L,S,S}(k,q,|\bq+\bk|;\eta,\eta',\eta') = - \frac{\bk \cdot \bq }{k^2} 
 u_2 u_a P^0\left(k\right) 
\left[ 1 - b_v \left( k, \eta \right) \right]  \nonumber\\ 
& & 
 \left\{   {\rm e}^{\eta'} P_{bc} \left( q; \eta', \eta' \right) \frac{ h_b^D-h_c^D }{ 2 }  
  - \int d s {\rm e}^s  \left[ {\bf h},{\bf \Omega} \right]_{mn} {\cal I}_{mnbc} \left( q; s, \eta', \eta' \right)
   \right\} \nonumber\\ 
& &   + {\rm O } \left( k^0 ;  \left( 1 - b_v \right)^2 \frac{\bk\cdot\bq}{k^2} \right)   \,, 
\label{consistency-equalt}
\eeqra 
with
\begin{eqnarray}
& & \!\!\!\!\!\!\!\!  \!\!\!\!\!\!\!\!  \!\!\!\!\!\!\!\!  \!\!\!\!\!\!\!\!\!\!\!\!\!\! 
 {\cal I}_{mnbc} \left( q; s, \eta', \eta' \right)  =   - \int d^3 p'   \frac{\bp' \cdot \bq}{2 q^2}  \int d s {\rm e}^s  \left[ {\bf h},{\bf \Omega} \right]_{mn}  \left\langle \frac{ \delta \left(  \vp_n \left(  \bp', s \right)   \vp_b \left( \bq, \eta' \right)   \vp_c \left( - \bq, \eta' \right) \right)  }{ \delta \vp_m \left(  \bp', s \right) } \right\rangle'  \nonumber\\ 
\end{eqnarray}
where we have used the relation \re{calT-def2}, and the fact that ${\cal Q}_{a}$ 
vanishes   when evaluated at equal time $\eta'=\eta''$. The  
$O\left(\left( 1 - b_v \right)^2 \frac{\bk\cdot\bq}{k^2} \right)$ corrections come from the $O(\Delta D^2)$ terms in eq.~\re{iDS}.

   Restricting our attention to the density contrast bispectrum, and making use of  the relations \re{multiplet}, this relation can be immediately cast in the form \re{consistency-delta} given in the introduction.  In \ref{isorole} we verify (at tree level) that, due to Galilean invariance, only internal line isocurvature modes contribute to the first two terms of eq.~\re{consistency-equalt}. 
   
Before closing this section, we come back to the fact that the contribution to the equal-time squeezed bispectrum that we have computed, eq.~\re{consistency-equalt}, is odd in $\bq$ or $\bk$, namely, it is of the form
\beq
\lim_{k\ll q}  B_{abc}^{L,S,S}(k,q,|\bq+\bk|;\eta,\eta',\eta') = \bk\cdot\bq \;f_{abc}(k,q;\eta,\eta'\eta')\,.
\label{sqodd}
\eeq
The reason for this angular behavior is in the shift defined by eqs.~\re{filrep} and \re{dispr}, namely, in the fact that the consistency relation takes into account the effect of a relative displacement between two different species due to velocity bias. On the other hand, any contribution to the squeezed bispectrum of the form \re{sqodd} can be removed by a proper shift of the fields, and, therefore, it can always be attributed to a relative displacement. Moreover, since  this effect is taken into account at a fully nonperturbative level, we conclude that  the corrections to eq.~\re{consistency-equalt}, indicated by $O(k^0)$, are either even in $\bq$ and $\bk$ or are suppressed by  $(1-b_v)^2$ factors \footnote{The third line of eq.~\re{Zshort}, that we have neglected, gives no contribution of the kind \re{sqodd} if the long wavelength fields are treated linearly. In any case, such contributions are suppressed by  $O(k^2/q^2)$ with respect to the ones included in the CR.}. This provides a powerful tool to single out such effects from the extra contributions to the bispectrum, on which we do not have nonperturbative control. For instance, for the equal time bispectrum, we could consider the combination
\beq
\lim_{k\ll q} \left( B_{112}^{L,S,S}(k,q,|\bq+\bk|;\eta,\eta',\eta') - B_{112}^{L,S,S}(k,q,|\bq-\bk|;\eta,\eta',\eta') \right)\,,
\label{oddcomb}
\eeq
which equals twice eq.~\re{consistency-equalt} with only $O((1-b_v)^2)$ corrections ({\it i.e.} no $O(k^0)$ ones).

Physically, a nontrivial relative displacement at equal times for the two species can be originated in two ways: either by a non-universal force for the two species, {\it i.e.}, by a violation of the equivalence principle,  by a statistical selection effect, or by non-universal initial conditions for the two velocity fields.   In the next section we discuss realizations of the first  and second possibilities. A physical situation were the third possibility is realized is studied in  \cite{Tseliakhovich:2010bj}, where the initial condition is an isocurvature mode in the   baryon-DM fluid.

\section{Two examples}
\label{examples}

In this Section we briefly review two contexts in which a velocity bias $b_v \neq 1$ emerges at large scales. In the first example, the nontrivial bias is due to the violation of the EP caused by a long range interaction between the DM and  the dark energy fields. The second example is a conventional DM scenario, and the bias is between the velocity of the DM and that of galactic halos.

\subsection{EP violation in the DM sector}
\label{EPDM}

In the first example \cite{Saracco:2009df}, a long range field $\phi$ possesses a direct coupling to cold DM, but not to baryons. Specifically, it is assumed that the mass of the DM particles depends on $\phi$. Moreover, $\phi$ may or may not play the role of Dark Energy. If the scalar field potential is negligible with respect to the cosmological constant the dynamics simplifies (as the time derivative of the vev of the field can be consistently set to zero), and therefore we will  make this assumption.   The resulting equation of motion for the scalar field is 
\begin{equation}
\Box \phi =  \beta \left( \phi \right) \left( T_{\rm DM} \right)^\mu_\mu \;\;,\;\; 
\beta \left( \phi \right) \equiv - \frac{d \; {\rm ln } \; m_{\rm DM} \left( \phi \right)}{d \phi} \;, 
\label{eom-phi}
\end{equation}
where $\Box$ is the d'Alambertian operator,   $ T_{\rm DM} $ is the DM  energy-momentum tensor, and $m_{\rm DM} \left( \phi \right)$ is the ($\phi-$dependent) DM mass. The equation (\ref{eom-phi}) is typical of scalar-tensor theories in the Einstein frame, where often the function $\beta$ is the same for all particles in order to respect the EP. In this model, universality is broken by assuming that only the DM mass depends on $\phi$. For simplicity, it is assumed that $\beta$ is constant. 

If $\beta=O(1)$,  DM clustering sources in comparable amount perturbations of the scalar field $\phi $ and the gravitational potential $\Phi$. The scalar field fluctuations in turn affect  the trajectories of the DM particles  \cite{Saracco:2009df}:
\begin{equation}
\partial_\tau \delta {v^j}_{\rm DM}  +{\cal H}  \delta {v^j}_{\rm DM} + \left( \delta {\bv}_{\rm CMD} \cdot {\bf \nabla} \right)  \delta {v^j}_{\rm DM} = - {\nabla^j } {\Phi} + \beta {\nabla^j } \phi \;, 
\label{euler-example1}
\end{equation}
where ${\cal H}$ is the Hubble rate in conformal time $\tau$. On the contrary, $\beta =0$ in the Euler equation for baryons. The (reduced) Planck mass has been set to $1$ in eq.   (\ref{euler-example1}). The Euler equations
for the DM and baryonic species are supplemented by continuity equations, the Poisson equation for $\Phi$, and a Poisson-like equation for $\phi$ coming from the newtonian limit of eq.~(\ref{eom-phi}). Using these last two equations to eliminate $\Phi$ and $\phi$, and  defining the multiplet (\ref{multiplet}), with the label $A$ for DM, and $B$ for baryons, one recovers a system of the  type (\ref{compact}), with 
\beqra
&&\!\!\!\!\!\!\!\!\!\!\!\!\!\! \!\!\! \!\!\! \Omega_{21} = -\frac{3}{2 f^2} \left((1-b)\Omega_B +\frac{\Omega_A}{b}\right)\,, \quad\Omega_{22}=\frac{3}{2 f^2} \left(\Omega_B +\frac{\Omega_A}{b}\right)\,,\quad\Omega_{23} = -\frac{3}{2f^2} \Omega_B\,\,,\nonumber\\
&&\qquad\qquad \Omega_{41}=-\frac{3}{2 f^2} \Omega_A\,,\qquad\qquad
\Omega_{43}=-\frac{3}{2 f^2} \Omega_B\,,
 \eeqra
 where $f$ and $b$ are the growth function and the (density and velocity) bias of the linear solution for this system, which has the form 
  $\vp_a(\bk,\tau) = \left\{ 1,f(\tau),b(\tau),b(\tau) f(\tau) \right\}_a \times \vp(\bk,\tau)$. $f$ and $b=b_v$ are the solutions of the system \cite{Saracco:2009df} 
 \beqra
 &&f'+\left(\frac{\cal{H}^\prime}{\cal{H}}+1\right)f+f^2-\frac{3}{2} \frac{\Omega_A}{b} -\frac{3}{2}\Omega_B=0\,,\nonumber\\
 && \frac{3}{2}\Omega_A (2 \beta^2 +1)+\frac{3}{2} b\, \Omega_B -\frac{3}{2}\frac{\Omega_A}{b}-\frac{3}{2}\Omega_B=0,
 \label{fb}
 \eeqra
 where primes denote derivatives with respect to conformal time. During matter domination, $f$ and $b$, as well as $\Omega_B=1-\Omega_A$ are constants, and so is the matrix $\Omega$. 
 
 It is immediate to see from the second equation above that the linear bias is unity for $\beta=0$, {\it i.e.} velocity (and density) bias emerges a consequence of the violation of the EP. 
As is well known, a bias also exists in the standard case $\beta = 0$, due to the different ``initial conditions'' for  the two species at decoupling. However, this bias tends to unity at low redshifts. For $\beta \neq 0$, even if this initial imbalance was zero, the velocity bias implied by eq.~\re{fb} is later established already at the linear level, and is further amplified by non-linear interactions  at smaller scales   \cite{Saracco:2009df}. 

\subsection{Halo velocity bias}
\label{HaloDM}

Let us move to the discussion of the second example, namely of the bias between the velocity of DM and of DM halos. The study of the formation of DM halos is the first step to understand galaxy clustering. Quite remarkably, also this problem can be successfully studied through the two-fluid system of equations given in  Sect.~\ref{invariance-tree}  \cite{2011MNRAS.416.1703E}. The first fluid is the DM, while the second is made of particles that eventually cluster to form halos of a given size, the so called  proto-halos. In N-body simulations, proto-halos are identified by tracing the positions of the particles that form a halo at $z=0$ back to the linear density field. One defines the position and the velocity of each proto-halo in the simulation from, respectively,  the center of mass and the mass weighted  velocity of these particle.  The fluid description can describe this system  up to scales greater than a few $\times$ the typical inter-halo separation, so that the discreteness of the proto-halo particles can be neglected. In this description, the proto-halo fluid does not contribute to the expansion of the universe and to the gravitational potential, since it is formed by particles already included in the DM fluid. Nonetheless, proto-halos evolve in the gravitational potential determined by the DM. Therefore, the system of DM and proto-halo obeys eq.~(\ref{compact}), where the matrix $\Omega$ is given by (\ref{Om-simple}) with $\Omega_A=\Omega_{DM} = 1$ and $\Omega_B = 0$. Ref.  \cite{2011MNRAS.416.1703E} computed the evolution of this system through the time renormalization group method of \cite{Pietroni08}, obtaining  halo-matter cross spectrum with a $5\%$ or better agreement with N-body simulations   up to $k \simeq 0.1 h {\rm Mpc}^{-1}$. 

Proto-halos form on the peaks of the DM distribution. Therefore, they sample special regions of the DM fluid, and this introduces a bias between the proto-halos and the DM \cite{Bardeen:1985tr}. Ref. \cite{Desjacques:2008jj} extended the computation of \cite{Bardeen:1985tr} of this bias by taking into account spatial derivatives of the linear density correlation. 

The velocity of peaks, which identify the proto-halos, is related to the DM density and velocity fields, linearly extrapolated to $z=0$, via
\beq
v^j_{pk}(\bx)=v^j_{DM,S}(\bx) -\frac{\sigma_0^2}{\sigma_1^2} \nabla^j \delta_{DM,S}(\bx)\,,
\label{hvb}
\eeq 
where, $\sigma_{0,1}$ are the first two momenta of the linear matter power spectrum $P \left( k , z  \right)$,
\begin{equation}
\sigma_n^2 \left( R_s, z \right)  \equiv \frac{1}{2 \pi^2} \int_0^\infty d k k^{2\left( n + 1 \right)} P \left( k , z  \right) 
W \left( k , R_s \right) \;,
\end{equation} 
computed at $z=0$, and the index ``$S$" on the DM density and velocity fields indicates a smoothing with the smoothing kernel $W \left( k , R_s \right)$, where the characteristic length $R_s$ corresponds to the lagrangian size of proto-halos of a given mass. 
In Fourier space, this results in a scale-dependent velocity bias    \cite{Desjacques:2008jj} 
\begin{equation}
b_v = \left( 1 - \frac{\sigma_0^2}{\sigma_1^2} k^2  \right) \, W \left( k R_s \right) \;.
\label{bv-ex2}
\end{equation}

Ref. \cite{Elia:2011ds} compared the result (\ref{bv-ex2}) against N-body simulations, obtaining a good agreement at large scales; for definiteness, let us consider the most massive halos in that study. From the numerical values given in Table 1 of  \cite{Elia:2011ds} we see that the analytic result (\ref{bv-ex2}) for the bias gives (at $z=0$)  
\begin{eqnarray}
&& \!\!\!\!\!\!\!\!\!\!\!\!\!\!\!\!  \!\!\!\!\!\!\!\!\!\!\!\!\!\!\!\! 
25 \leq \frac{M_{\rm halo}}{10^{12} h^{-1} M_\odot} \leq 40 \;: \;\;\;\; b_v \left( k = 0.05 h {\rm Mpc}^{-1} \right) \simeq 0.95  
\;\;,\;\;  b_v \left( k = 0.1 h {\rm Mpc}^{-1} \right) \simeq 0.82 \;, \nonumber\\ 
&&  \!\!\!\!\!\!\!\!\!\!\!\!\!\!\!\!  \!\!\!\!\!\!\!\!\!\!\!\!\!\!\!\! 
40 \leq \frac{M_{\rm halo}}{10^{12} h^{-1} M_\odot} \leq 100 \;: \;\;\;\; b_v \left( k = 0.05 h {\rm Mpc}^{-1} \right) \simeq 0.93 
\;\;,\;\;  b_v \left( k = 0.1 h {\rm Mpc}^{-1} \right) \simeq 0.73 \;,  \nonumber\\ 
\end{eqnarray} 
and we see that the bias can be substantially different from one already at those large scales. As visible in Figure 4 of  \cite{Elia:2011ds}, the N-body simulation provides slightly smaller values for the bias, which are in good agreement with those of the analytic computation.

Notice that, as discussed at the end of the previous section, the only terms proportional to $\bq\cdot\bk$ come from velocity bias, so that,  even if  the bias in eq.~\re{bv-ex2} is proportional to $k^2$, it can be distinguished from the higher order bias-independent effects, neglected in the derivation of the CR, by considering odd combinations of bispectra such as eq.~\re{oddcomb}. 

The halo velocity bias \re{hvb} is proportional to $\frac{\sigma_0^2}{\sigma_1^2}\nabla^j \delta_{DM}$, whereas that induced by the EP violation of the previous example, eq.~\re{euler-example1}, goes as $\beta\, \nabla^j \phi \propto \beta^2 \frac{\nabla^j}{\nabla^2} \delta_{DM}$. This would result in contributions to the CR  proportional to $\bq\cdot\bk\, \sigma_0^2/\sigma_1^2$ and $\beta^2\, \bq\cdot\bk/k^2$, respectively.  The consideration of equal-time CR as in eq.~\re{consistency-equalt} can open interesting perspectives to the measurement of large scale velocity bias. Taking into account only the angular dependence would not allow a separation between the effect due to EP violation and that due to a velocity bias of statistical origin, or induced by different initial conditions, as in \cite{Tseliakhovich:2010bj}. However, by considering different $(k,q)$ pairs these two effects can be in principle disentangled.

\section{Inclusion of additional effects}
\label{bssa}

The CR derived in this paper involve correlators and cross-correlators between density and velocity fields for which one can write a continuity equation and a Euler equation, possibly including non-gravitational long range interactions, as discussed in Sect.~\ref{EPDM}. These equations are expressed in compact form in eq.~\re{compact}. To assess the range of applicability of the CR, one needs to study the impact of other physical effect, not included in \re{compact},  on the density and velocity fields of a given species. 

We can distinguish three different effects (a) multistreaming, (b) baryon-photon interactions, 
(c) formation, merging, disruption.

We now study them separately. From an intuitive point of  view, it is clear that a small scale effect that is  insensitive to a large scale boost (namely, that is insensitive to a shift like eq.~\re{filrep} or \re{shiftmom}) will induce extra terms in the evolution equations that ultimately do not modify the CR. This is the case for the two effects (a) and (b) just mentioned.

Let us first show  that the effect (a) does not modify our CR.  We start by discussing explicitly the effect of multi streaming on the CR for a pure DM system. The starting point is the Vlasov equation for the DM distribution function $f(\bx, \bp, \tau)$ which gives the full description of the system in phase space
\beq
\left(\frac{\partial}{\partial \tau}+ \frac{p^i}{ a m}\frac{\partial}{\partial x^i}- am\nabla^i_x \phi(\bx,\tau) \frac{\partial}{\partial p^i}\right)\,f(\bx,\bp,\tau)=0\,,
\label{vlasov}
\eeq
where
\beq
\frac{d x^i}{d\tau} = \frac{p^i}{am}\,,\qquad\qquad \frac{d p^i}{d\tau} = - am\nabla^i_x \phi\,.
\eeq
In order to work in configuration space, we define moments of the coarse-grained distribution function as usual,
\beqra
&&  n(\bx,\tau) = \int d^3 p  f(\bx,\bp,\tau) \,,\nonumber\\
&& v^i(\bx,\tau)= \frac{1}{ n(\bx,\tau)} \int d^3 p \frac{p^i}{am}  f(\bx,\bp,\tau) \,,\nonumber\\
&& \sigma^{ij}(\bx,\tau)= \frac{1}{ n(\bx,\tau)} \int d^3 p \frac{p^i}{am} \frac{p^j}{am} f(\bx,\bp,\tau) - v^i(\bx,\tau) v^j(\bx,\tau) \nonumber\\
&& \cdots\,.
\label{n-v-sigma-bar}
\eeqra
If one is interested in following the density and velocity fields only, the Vlasov equation, eq.~\re{vlasov}, is then completely equivalent to the system (omitting the $(\bx,\tau)$ dependence of the various quantities)
\beqra
&&\frac{\partial}{\partial \tau}  \delta +\frac{\partial}{\partial x^i} \left((1+ \delta )\, v^i  \right) =0\,, \label{continuity}\\
&&\frac{\partial}{\partial \tau}  v^i+{\cal H} v^i +  v^k\frac{\partial}{\partial x^k} v^i  = -\nabla_x^i \phi -J_\sigma^i\,,\label{euler}
\eeqra
where the density fluctuation,
\beq
\delta(\bx,\tau) = \frac{n(\bx,\tau)}{n_0}-1,
\label{delta-bar}
\eeq
is related to the potential by the Poisson equation,
\beq
\nabla^2 \phi = \frac{3}{2}{\cal H}^2\Omega_m \,\delta\,.
\label{Poisson}
\eeq
The source term at the RHS of the Euler equation is given by 
\beqra
&& J_\sigma^i(\bx,\tau) \equiv\frac{1}{1+\delta} \frac{\partial}{\partial x^k}\left((1+ \delta) \sigma^{ik}\right), 
\label{Jsig}
\eeqra
and contains all the information on higher moments of the distribution function.
The single stream approximation amounts to neglecting $J_\sigma^i$.  In other words, all the effects of short scale multi streaming are transmitted to the velocity and density fields by this source term.
Since it is manifestly invariant under uniform velocity boosts of the form $p^i/(am) \to p^i/(am) + \dot d^i(\tau)$ (see eq.~\re{n-v-sigma-bar})  its inclusion plays no role in the derivation of the Ward identities or of the CR. For instance, the transformation of the source would add no terms to eqs.~\re{dZ1} or \re{Zshort}, and therefore the conclusions of the previous sections would be unaltered. 

In the previous sections we have neglected  the vorticity component of the velocity field, ${\bf \omega}\equiv \nabla\times\bv$. However, by looking at its impact on the equations of motion for $\delta$ and $\theta$,  it is easy to realize that a variation of  ${\bf \omega}$ under a uniform boost does not give nontrivial contributions to the variation of the non-linear terms. 

We therefore showed that the effect (a) does not modify the CR in the case of a single DM fluid. This conclusion can be immediately generalized  to the case of  multi DM species, or to a system of DM and baryons:  each Euler equation for the different velocity fields gets an independent source, of the same structure as eq.~ \re{Jsig}, which will be invariant under independent uniform shifts of the velocity  fields. The presence of extra non-gravitational and EP violating long range forces, such as that discussed in Sect.~\ref{EPDM}, would not modify the argument above, as it impacts the first term at the RHS of \re{euler}  but not the source term. 

Let us now comment on the effect (b). We are referring to the local Quantum Electro Dynamics (QED) interactions between photons and baryons. The baryon-photon interactions    are taken into account by a collisional term on the RHS of the Vlasov (Boltzmann) equation. This term is also independent on any boost acting on the same way on baryons and photons, and so also the effect (b) does not modify the CR.

Coming to the discussion of the effect (c), our study does not include the possibility of considering non-conserved objects such as galaxies or halos undergoing the process of formation, merging or disruption, which modifies the equations for that population (irrespectively of whether the EP is or is not violated) in a way that  could possibly   prevent the possibility of deriving CR-like relations. Therefore the process (c) is beyond  the scope of our work, and - to our knowledge - of the existing CR literature. The  CR we have derived hold for systems for which it is possible to write a continuity equation, as for instance the  proto-halos discussed in  Sect. \ref{HaloDM}.

\section{Conclusions}
\label{conclus} 
The main result of this paper is eq.~\re{nexwardbias}, which generalizes the consistency relation found in refs.\cite{Peloso:2013zw,Kehagias:2013yd} to the multi-species case. In presence of large-scale velocity bias between the different species, the consistency relation gets extra terms including products of nonlinear PS's, of nonlinear propagators and PS, and a nonlinear trispectrum.  
The leading ${\rm O } \left( \frac{\bk \cdot \bq}{k^2} \right)$ terms, where $k$ and $q$ are the soft mode and hard modes, respectively, are exact non-perturbatively.

We stress that, as for the CR obtained in \cite{Peloso:2013zw,Kehagias:2013yd}, also these generalized ones are valid beyond the single stream approximation, as we discuss in Sect.~\ref{bssa}, and even including extra long-range forces.

We have discussed the relation between WI and CR (In particular, see  \ref{app:WICR}). The former reflect the symmetry property of the action, or of the equations of motion, and represent non-perturbative relations between 1PI or amputated Green functions. There relations provide powerful constraints on the non-linear structure of the theory, analogously to what happens in quantum field theory, where they relate different terms of the effective action of the fully renormalized theory. To pass from WI to CR we need to add information on the structure of the solutions of the equations of motion at long wavelengths. The CR found in this paper, as we have seen, are sensitive to a velocity bias. We have discussed two possible scenarios in which such a bias can occur: a new long range scalar force in the DM sector, and the velocity bias of DM halos with respect to the underlying DM field. From an observational point of view, the $b_v-1$ terms in eq.~\re{nexwardbias} are more promising than the $b_v+1$ ones, as they do not necessarily vanish when the correlators are evaluated at equal times. In particular, these terms could be tested cross-correlating different galaxy populations, or different surveys, as galaxy surveys and weak lensing ones. As we mentioned after eq.~\re{eq45}, the extraction of the bias parameter $b_v$  from these measurements would require the computation of some of the $b_v-1$ terms by either N-body simulations or some semi analytical approximation scheme. 
We hope to come back to this in a separate publication.

\section*{Acknowledgments}

We thank the authors of  ref.~\cite{Creminelli:2013mca} for valuable discussions.  We thank an anonymous referee for  useful comments on an earlier version of this manuscript. M. Pietroni acknowledges partial support from the European Union FP7 ITN INVISIBLES (Marie Curie Actions, PITN- GA-2011- 289442) and by the European Programme PITN-GA-2009-237920 (UNILHC ).  M. Peloso  acknowledges partial support from the DOE grant DE-FG02-94ER-40823 at the University of Minnesota.   

\appendix

\section{From WI to CR}
\label{app:WICR}

In Section \ref{ward} we have studied how to obtain WI associated with transformations of the fields of the form of a generalized Galilean transformations. In Section \re{consistency} we have instead shown how to obtain  CR from a change of variable in the generating functional and from the solutions of the equations of motion at large scales. In this Appendix we show the link between WI and CR, namely how the latter can be directly obtained from the former, and from the large scale solutions. We start by taking the double derivative of the expression \re{WIconW}. We obtain 
\beqra
&& \lim_{k\to 0} d_n\;\partial_\eta \bigg[e^{-\eta} (-\delta_{nm}\partial_\eta +\Omega_{mn}) B^{\chi\vp\vp}_{mbc} (k,q,|\bk+\bq|;\eta,\eta',\eta'')\bigg]\nonumber\\
&&\qquad\qquad = \frac{\bk\cdot\bq}{k^2} P_{bc}(q;\eta',\eta'') \bigg(\delta_D(\eta-\eta') - \delta_D(\eta-\eta'')\bigg) +O(k^0)\,,
\label{wint1}
\eeqra
where we have used the relation \re{PSdef} for the PS, as well as the relation
\beq
 \!\!\!\!\! \!\!\!\!\!\!\! \!\!\!\!\!\!\! \!\!\!\!\!\!\! \!\!\!\!\!\!\!\left.\frac{\delta^3 W}{\delta K_l(\bk,\eta) \delta J_m(\bk',\eta')\delta J_n(\bk'',\eta'') } \right|_{J_m=K_n=0} = -i  \delta_D(\bk+\bk'+\bk'') B_{lmn}^{\chi\vp\vp}(k;k',k'';\eta,\etap,\eta'')\,,
\eeq
for the bispectrum. 
To obtain eq.~\re{wint1}  we have used the fact that the equation
\begin{equation}
 \int d^3 k \delta_D \left( \bk \right) \bk \cdot \bw F \left( \bk,\bq \right) =   \bq\cdot \bw \; G \left(\bq \right) \;,
\end{equation}
is formally solved by   
\begin{equation}
\lim_{k\rightarrow 0}  F \left( \bk,\bq \right) = \frac{\bk \cdot \bq}{k^2}  G \left(\bq \right)+ {\rm O } \left( k^0 \right) \,,
\label{formal-sol}
\end{equation}
and that $\bw(\eta)$ can be chosen arbitrarily.

\begin{figure}[ht!]
\centerline{
\includegraphics[width=1.\textwidth,angle=0]{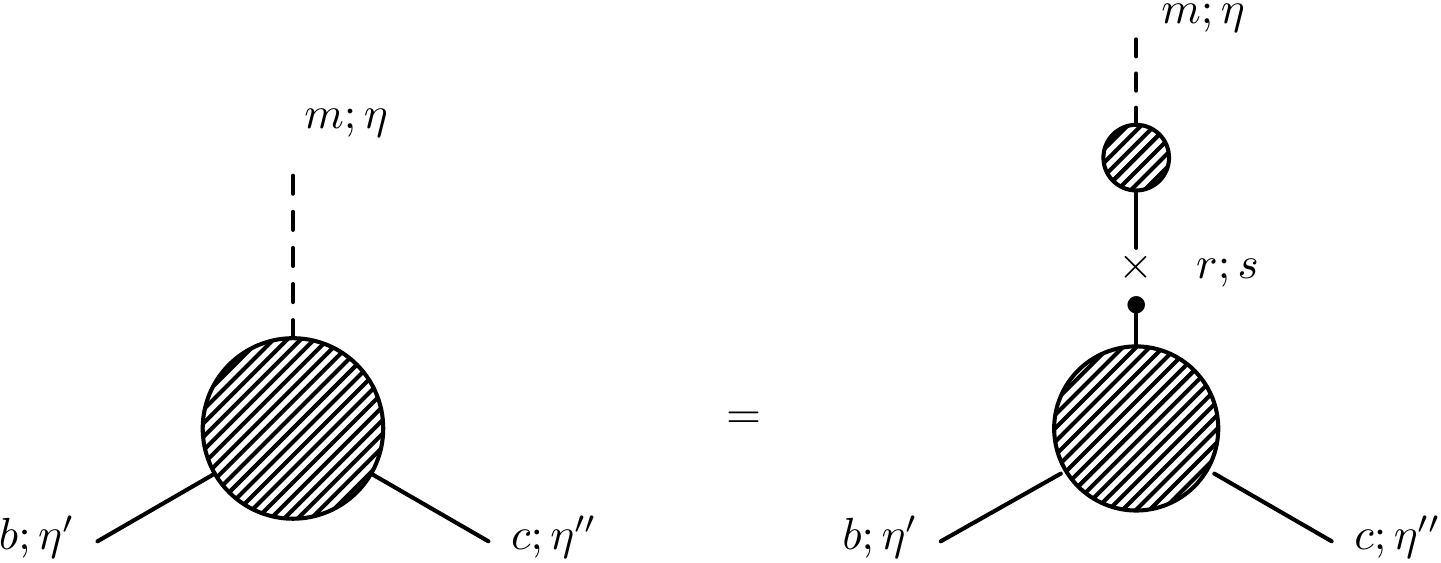}
}
\caption{Diagrammatic expression of the first equality in (\ref{ampute-connect}).
The two labels on each external lines and on the vertex denote the field and the time, respectively. 
}
\label{ampute}
\end{figure}

The three-point function  $B_{mbc}^{\chi\vp\vp}$  appearing in the above relation (\ref{wint1}) and     can be written as 
\beqra
&& B_{mbc}^{\chi\vp\vp}(k,q,p,\eta,\eta',\eta'') \equiv \int ds\; G_{rm}(k;s,\eta)\, B_{rbc}^{[\vp]\vp\vp}(k,q,p,s,\eta',\eta'') \,,\nonumber\\
&&\qquad\longrightarrow   \int ds\; g_{rm}(s-\eta)\, B_{rbc}^{[\vp]\vp\vp}(k,q,p,s,\eta',\eta'')  \qquad(\mathrm{for}\; \;k\to0)\,,
\label{ampute-connect}
\eeqra
where $B_{rbc}^{[\vp]\vp\vp}(k,q,p,s,\eta',\eta'') $ is the connected three-point function in which the leg connecting to the first $\vp$ ending has been amputated (see Fig.~\ref{ampute})\footnote{Analogously, the four-point function in \re{trisp} can be expressed as $T_{lmno}^{\chi\vp\vp\vp}(\bk,\bk',\bk'',\bk''';\eta,\etap,\eta'',\eta''') = \int ds\; G_{rl}(k;s,\eta) T_{rmno}^{[\vp]\vp\vp\vp}(\bk,\bk',\bk'',\bk''';s,\etap,\eta'',\eta''')$, where $T_{rmno}^{[\vp]\vp\vp\vp}$ is the trispectrum $\langle\vp_r \vp_m\vp_n\vp_o\rangle$ in which the leg connecting to the ending labeled by the index $``r"$ has been amputated.}.

Using
\beq
(-\delta_{nm}\partial_\eta+\Omega_{mn})g_{rm}(s-\eta) = (\delta_{nm}\partial_s+\Omega_{mn})g_{rm}(s-\eta) = \delta_{rn}\delta(s-\eta)\,,
\eeq
where we have used the property of the linear propagator $ [{\bf g}(\eta), {\bf \Omega}]=0$, we see that the effect of the operator multiplying $B^{\chi\vp\vp}$ in eq.~\re{wint1}   is to cut (to ``amputate'')  the propagator, that is
\beq
 \!\!\!\!\!\!\!\!\!\!\!\!\!\!(-\delta_{nm}\partial_\eta +\Omega_{mn}) B^{\chi\vp\vp}_{mbc} (k,q,|\bk+\bq|;\eta,\eta',\eta'') = B^{[\vp]\vp\vp}_{nbc} (k,q,|\bk+\bq|;\eta,\eta',\eta'')\,,
\eeq
so that the object appearing at the first line of  eq.~\re{wint1}   reads
\beqra
&& \!\!\!\!\!\!\!\!\!\!\!\!\!\!\partial_\eta \bigg[e^{-\eta} (-\delta_{nm}\partial_\eta +\Omega_{mn}) B^{\chi\vp\vp}_{mbc} (k,q,|\bk+\bq|;\eta,\eta',\eta'')\bigg] \nonumber\\
&&\qquad\qquad= - e^{-\eta}(1-\partial_\eta) B^{[\vp]\vp\vp}_{nbc} (k,q,|\bk+\bq|;\eta,\eta',\eta'')\,,
\label{bispamp}
\eeqra
multiplying \re{wint1}   by $e^\eta$ and integrating in $\eta$ we obtain
\beq
 \!\!\!\!\!  \!\!\!\!\!  \!\!\!\!\!  \!\!\!\!\!  \!\!\!\!\!  \!\!\!\!\! 
   \int ds \; d_n B^{[\vp]\vp\vp}_{nbc} (k,q,|\bk+\bq|;s,\eta',\eta'') =-\frac{\bk\cdot\bq}{k^2} P_{bc}(q;\eta',\eta'') \bigg(e^{\eta''} -e^{\eta'}\bigg)+O(k^0)\,,  
 \label{WI1}
\eeq
where we have relabeled the integration time from $\eta$ to $s$. As we shall see, this identity is of a suitable form 
 to derive the CR that we had already obtained in the main text. Before doing so, however, we need to derive an identity from transformations in which the two species transform differently.  Let us replace eq. \re{infsplitboost} with 
\beqra
&& \delta\varphi_n \left( \bk , \eta \right) =  i \bk\cdot  \Delta \bw(\eta) h_{nm} \;
\varphi_m \left( \bk , \eta \right)   + i  \, {\rm e}^{-\eta}  \; \bk\cdot \partial_\eta  \Delta \bw(\eta)h_{nm}  \;  d_m \,\delta_D(\bk)\,,\nonumber\\
&& \delta\chi_n \left( \bk , \eta \right) =  i \bk\cdot  \Delta \bw(\eta) h_{nm}\;
\chi_m \left( \bk , \eta \right) \,, 
\label{infsplitboost2}
\eeqra
in which the two species transform with an opposite shift.  Starting from this transformation, and following the exact same steps that led from  eq. \re{infsplitboost} to eq. \re{WI1}, we obtain 
\beqra
&&\!\!\!\!\! \!\!\!\!\!  \!\!\!\!\!  \!\!\!\!\! \int ds \; h_{nr} d_r B^{[\vp]\vp\vp}_{nbc} (k,q,|\bk+\bq|;s,\eta',\eta'') = \nonumber\\
&&\!\!\!\!\! \!\!\!\!\!  \!\!\!\!\!  \!\!\!\!\! =- \frac{\bk\cdot\bq}{k^2}  \bigg(h_{bl} P_{lc}(q;\eta',\eta'') \,\e^{\eta'} -h_{cl} P_{lb}(q;\eta'',\eta')\, \e^{\eta''}\bigg)\nonumber\\
&&\!\!\!\!\! \!\!\!\!\!  \!\!\!\!\!  \!\!\!\!\! - \frac{\bk\cdot\bq}{k^2} \;[{\bf h},{\bf \Omega}]_{mn} \int ds \,\e^\eta \bigg(G_{bm}(q;\eta',s) P_{nc}(q;s,\eta'') - G_{cm}(q;\eta'',s) P_{nb}(q;s,\eta') \bigg)\nonumber\\
&& \!\!\!\!\! \!\!\!\!\!  \!\!\!\!\!  \!\!\!\!\! + [{\bf h},{\bf \Omega}]_{mn}\int ds\,\e^s\, \int d^3\bp \; \frac{\bk\cdot\bp}{k^2} T^{\chi\vp\vp\vp}_{mnbc}(\bp,-\bp,\bq,-\bq;s,s,\eta',\eta'')+O(k^0)\,.
\label{WI2}
\eeqra

We can now obtain the CR for the full bispectrum. To pass from the  relations \re{WI1} and \re{WI2} involving the amputated bispectrum, to relations involving the full bispectrum, we have to attach a soft PS to the amputated leg, and then integrate over the time corresponding to the insertion point. Indeed,  the contributions to the bispectrum in which the  external soft leg is a propagator instead of a PS are subdominant (that is, not enhanced by the $\bq\cdot \bk/k^2$ term). This is due to the fact that the vertex inside the diagram in which the $\chi$ line of the propagator is attached vanishes in the $k \rightarrow 0$ limit. In writing the soft PS, we will use the solution for the linearized modes. This will explicitly show that the CR are equivalent to the WI - that  gave eqs. \re{WI2} and \re{WI2} - plus the solution for the large scale modes. 

The linear PS has the form
\beq
P_{mn}(k;\eta,\eta') \to P^0(k) u_m(\eta) u_n(\eta')\,,
\label{lps}
\eeq
where $P^0(k)$ is the initial bispectrum and $u_m(\eta)=g_{mn}(\eta)u_n^{in}$, with $u_n^{in}$ encoding the structure of the initial conditions for the fields, namely $$\langle \vp^{in}_m(\bk) \vp^{in}_n(\bk')\rangle = \delta_D(\bk+\bk') P^0(k) u_m^{in}u_n^{in}\,.$$
The bispectrum in the soft $k$  limit is therefore
\beqra
&&B^{\vp\vp\vp}_{abc} (k,q,p;\eta,\eta',\eta'') =  P^0(k) u_a(\eta) \int ds \;u_n(s) B^{[\vp]\vp\vp}_{nbc} (k,q,p;s,\eta',\eta'') \,,\nonumber\\
&&\qquad\qquad\qquad \qquad\qquad\qquad \qquad\qquad\qquad \qquad\qquad\qquad (k\to  0)
\label{bisp}
\eeqra
where the leading contributions to the sum over $n$ in the soft limit are those corresponding to $n=2,4$. 
If the initial conditions are on the growing mode, then $u_n(\eta)$ is constant  in time. We can decompose it as 
\beq
u_n=\frac{d_n+h_{nm}d_m}{2} u_2 +\frac{d_n-h_{nm}d_m}{2}  u_4+\cdots\,,
\label{compo0}
\eeq
where the dots indicate the $n=1,3$ components which, once contracted with the bispectrum give subdominant contributions in the $k\to 0$ limit.
Therefore, in terms of the linear velocity bias introduced in \re{d3}, we can write 
\beq
 u_n B^{[\vp]\vp\vp}_{nbc} =  d_n B^{[\vp]\vp\vp}_{nbc}\frac{1+b_v}{2} u_2 + h_{nm}d_m B^{[\vp]\vp\vp}_{nbc}\frac{1-b_v}{2} u_2\,,
\label{comb}
\eeq

It is now immediate to verify that, taking the combination of eqs.~\re{WI1} and \re{WI2} according to \re{comb}, we reobtain the CR \re{nexwardbias2} of the main text.  More precisely, this derivation reproduces eq.  \re{nexwardbias2}, apart from the ${\cal Q}_{bc}$ term. This term arised from  contributions of order $(\Delta{\bf D})^2$ or $(D(\eta')-D(\eta''))\Delta{\bf D}$     in the transformation \re{shiftmom}, cfr. eqs. \re{e3} and \re{longcont-term2}.  To reproduce such a term in the present derivation we would need to 
consider variations of the partition function $Z$ beyond the linearized order in $ \bw(\eta) $ and in  $\Delta \bw(\eta)$ 
that led to the identities \re{WI1} and \re{WI2}, respectively. For brevity, we do not perform this computation here.

\section{Explicit check at tree level}
\label{treetest}

\begin{figure}[ht!]
\centerline{
\includegraphics[width=1.\textwidth,angle=0]{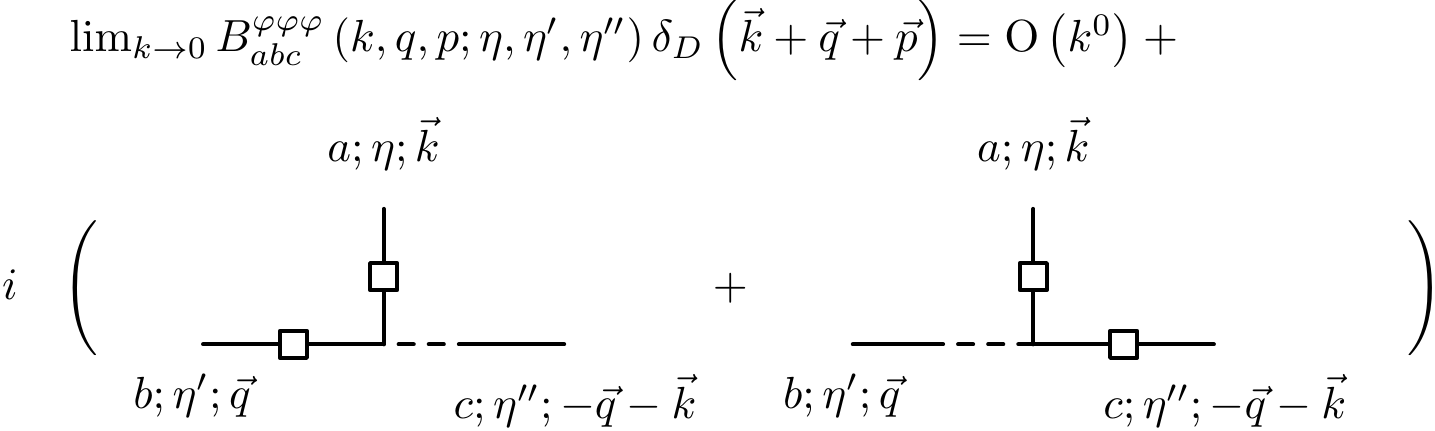}
}
\caption{Diagrammatic expression of the LHS of (\ref{nexwardbias2}) at tree level. The  labels on each external line and on the  denote the filed, the time, and the momentum,  respectively. Our convention is that all momenta point towards the internal vertex.
}
\label{tree}
\end{figure}

In this Appendix we verify the CR (\ref{nexwardbias2}) - or, equivalently, eq.  (\ref{nexwardbias}) with the linear mode $k$ on the growing mode  - 
 at tree level. The diagrammatic expression for the LHS is given in Figure \ref{tree}. The solid line with a box denotes a linear power spectrum (the tree level expression of 
(\ref{PSdef})), while the solid/dashed line  denotes a linear propagator (the tree level expression of (\ref{propdef})). 

We note that the figure does not include  a diagram with the propagator on the $\vp_a \left( k \right)$  line, since the vertex vanishes when attached to a propagator of vanishing momentum.  The expression in the Figure evaluates to 
\begin{eqnarray}
&&\!\!\!\! \!\!\!\! \!\!\!\! \!\!\!\! \!\!\!\! B^{\vp\vp\vp}_{abc,{\rm tree}} \left( k,q,p;\eta,\eta',\eta'' \right) \delta_D \left( \bk+\bq+\bp \right)  =  \nonumber\\
&& \!\!\!\! \!\!\!\! \!\!\!\! \!\!\!\! +2 \int^{\eta''} d s  \;u_b P^{(0)} \left( q \right) u_{b'} \;
 e^s \gamma_{c'a'b'} \left( - \vec{q}-\vec{k},\vec{k},\vec{q} \right) g_{cc'} \left( \eta'' - s\right)  u_{a'} P^{(0)} \left( k \right)  u_a  \nonumber\\
&& \!\!\!\! \!\!\!\! \!\!\!\! \!\!\!\! 
+2 \int^{\eta'} d s \;
  g_{bb'} \left( \eta' - s\right) 
 e^s \gamma_{b'a'c'} \left( \vec{q}, \vec{k}, - \vec{q}-\vec{k} \right)
  u_{c'} P^{(0)} \left( \vert \vec{q} + \vec{k} \vert \right)  u_c 
u_{a'} P^{(0)} \left( k \right) u_a \nonumber\\
&&\qquad\qquad \qquad\qquad \qquad\qquad \qquad\qquad + {\rm O } \left( k^0 \right)\,.
  \label{check1}
 \end{eqnarray}
 In the $k\rightarrow0$ limit, the expressions (\ref{vertex}) for the vertex reduce to   
\begin{eqnarray}
\!\!\!\! \!\!\!\! \!\!\!\! \!\!\!\! \!\!\!\! \lim_{p\ll k} \gamma_{abc} \left( \bk , \bp , \bq \right) &=& \frac{\bp \cdot \bq}{2 p^2}  \delta_D \left( \bk + \bq \right) 
\left[ \delta_{b2} \left( \delta_{a1} \delta_{c1} +  \delta_{a2} \delta_{c2}  \right) +  \delta_{b4} \left( \delta_{a3} \delta_{c3} +  \delta_{a4} \delta_{c4}  \right)  \right] \nonumber\\
&&+ {\rm O } \left( p^0 \right) \nonumber\\
 & = & \frac{\bp \cdot \bq}{4 p^2}  \delta_D \left( \bk + \bq \right) 
\left[ \delta_{ac} \delta_{bd} +  h_{ac} h_{bd}  \right] d_d  + {\rm O } \left( p^0 \right) 
\end{eqnarray}
Using this expression, and the identities 
\begin{equation}
u_a \, d_a = \left( 1 + b_v \right) u_2 \;\;\;,\;\;\; u_a \, h_{ab} d_b = \left( 1 - b_v \right) u_2
\end{equation}
the expression (\ref{check1}) reduces to 
\begin{eqnarray}
\!\!\!\! \!\!\!\! \!\!\!\! \!\!\!\! \!\!\!\! && \lim_{k\ll q} B^{\vp\vp\vp}_{abc,{\rm tree}} \left( k,q,p;\eta,\eta',\eta'' \right) = \nonumber\\
\!\!\!\! \!\!\!\! \!\!\!\! \!\!\!\! \!\!\!\! && P^{(0)} \left( q \right) P^{(0)} \left( k \right) \frac{\bk \cdot \bq}{2 k^2} 
u_2 u_a u_b   \left[ \int^{\eta''} d s \;{\rm e}^s g_{cd} \left( \eta'' - s \right) \right]  
 \left[ \left( 1 + b_v \right) \delta_{de} +  \left( 1 - b_v \right) h_{de}  \right] u_e 
\nonumber\\ 
 \!\!\!\! \!\!\!\! \!\!\!\! \!\!\!\! \!\!\!\! &  & -  \left( \eta'' \rightarrow \eta' ; b \leftrightarrow c \right) + {\rm O } \left( k^0 \right) 
\label{treeB}
\end{eqnarray}

We are interested in the case in which the external fields are along a growing mode. Namely they are
null eigenmodes of (\ref{BIGOMEGA}) and they satisfy $ g_{cd} \left( \eta'' - s \right) u_d = u_c$. 
This simplifies the term $\propto 1 + b_v$ in the above relation. In the other term, we rewrite 
\begin{eqnarray}
 \!\!\!\! \!\!\!\! \!\!\!\! \!\!\!\! \!\!\!\!\int d s {\rm e}^s g_{cd} \left( \eta'' - s \right) h_{de} u_e & = &  \int d s \left[ \partial_s {\rm e}^s \right] g_{cd} \left( \eta'' - s \right) h_{de} u_e \nonumber\\
&=& \int d s {\rm e}^s \partial_{\eta''} g_{cd } \left( \eta'' - s \right) h_{de} u_e 
 \nonumber\\ 
 \!\!\!\! \!\!\!\! \!\!\!\! \!\!\!\! \!\!\!\!& = & \int d s {\rm e}^s  \left[ \left(  \delta_{gd} \partial_{\eta''} + \Omega_{gd} \right) - \Omega_{gd} \right]  g_{cg } \left( \eta'' - s \right) h_{de} u_e  \nonumber\\
 \!\!\!\! \!\!\!\! \!\!\!\! \!\!\!\! \!\!\!\!& = & \int d s {\rm e}^s \left[ \delta_{cd} \delta_D \left( \eta'' - s \right)  - g_{cg} \left( \eta'' - s \right) \Omega_{gd} \right] h_{de} u_e \nonumber\\ 
 \!\!\!\! \!\!\!\! \!\!\!\! \!\!\!\! \!\!\!\!& = & {\rm e}^{\eta''} h_{ce} u_e - \int d s {\rm e}^s g_{cg} \left( \eta'' - s \right) \left[ \Omega , h \right]_{ge} u_e \;\;,
\end{eqnarray} 
where in the last expression we have used the fact that the growing mode $u_e$ is a zero mode of $\Omega$. 

We therefore find 
\begin{eqnarray}
&& \!\!\!\!\!\!\!\!\!\!\!\!\!\!\!\!  \!\!\!\!\!\!\!\!\!\!\!\!\!\!\!\! 
 \lim_{k\ll q} B^{\vp\vp\vp}_{abc,{\rm tree}} \left( k,q,p;\eta,\eta'',\eta' \right) = 
 P^{(0)} \left( q \right) P^{(0)} \left( k \right) \frac{\bk \cdot \bq}{2 k^2}  \nonumber\\ 
 \!\!\!\!\!\!\!\!\!\!\!\!\!\!\!\!    \!\!\!\!\!\!\!\!\!\!\!\!\!\!\!\!  \!\!\!\!\!\!\!\!\!\!\!\!\!\!\!\!&&  \!\!\!\!\!\!\!\!\!\!\!\!\!\!\!\!  \!\!\!\!\!\!\!\!\!\!\!\!\!\!\!\! \times u_s u_a \left\{ \left( 1 + b \right) u_b u_c  {\rm e}^{\eta'' } 
 + \left( 1 - b \right)  u_b \left[ h_{cd} u_d {\rm e}^{\eta''} + \int^{\eta''} d s {\rm e}^s g_{cd} \left( \eta'' - s ; q  \right) \left[ h, \Omega  \right]_{de} u_e \right] \right\} \nonumber\\ 
  \!\!\!\!\!\!\!\!\!\!\!\!\!\!\!\! \!\!\!\!\!\!\!\!\!\!\!\!\!\!\!\! &  & -  \left( \eta'' \rightarrow \eta' ; b \leftrightarrow c \right) + {\rm O } \left( k^0 \right) \;.
\label{lhs}
\end{eqnarray}
It is immediate to verify that this expression coincides with the tree level expansion of the RHS of (\ref{nexwardbias2}) (we note that the last term in (\ref{nexwardbias2}) vanishes at tree level, as the $4$-point function 
$T^{\chi\vp\vp\vp}$ can enter in a diagram for the bispectrum only through a loop; the same is true for the quantitiy  ${\cal Q}_{bc}$). We therefore have explicitly  verified the relation (\ref{nexwardbias2}) at tree level.

\section{Contribution of the isocurvature modes to the equal time squeezed bispectrum}
\label{isorole}

The equal time squeezed bispectrum $\lim_{k\ll q} B \left( \bq , \bk , - \bq - \bk ; \eta, \eta' , \eta'' = \eta' \right)$
measures how the power spectrum of the hard mode $P \left( q ; \eta' \right)$ is modulated by the presence of a soft mode of momentum $\bk$. The term enhanced by $\frac{q}{k}$ in the relation  (\ref{nexwardbias})  is proportional to the large scale velocity modes $u_2$ and $u_4$. Due to Galilean invariance, short modes are not affected by a   large scale adiabatic velocity mode. Therefore we should expect that the ${\rm O } \left(  \frac{q}{k} \right)$ contribution from the adiabatic mode vanishes at equal time $\eta'' = \eta'$.  The purpose of this Subsection is to verify this explicitly at tree level. The matrix (\ref{BIGOMEGA}) has the eigenvalues
\begin{eqnarray}
& & \!\!\!\!\!\!\!\!\!\!\!\!\!\!\!\! \!\!\!\!\!\!\!\!\!\!\!\!\!\!\!\! 
\lambda_{1,2}  =  \frac{1}{2} \left[ 1 + \Omega_{22} \mp \sqrt{1-2\Omega_{21}-2 \Omega_{22} + \Omega_{22}^2-2\Omega_{43} + 2 \sqrt{\left( \Omega_{21}-\Omega_{43} \right)^2 + 4 \Omega_{23} \Omega_{41}}} \right] \;,  \nonumber\\ 
& & \!\!\!\!\!\!\!\!\!\!\!\!\!\!\!\! \!\!\!\!\!\!\!\!\!\!\!\!\!\!\!\! 
\lambda_{3,4}  =  \frac{1}{2} \left[ 1 + \Omega_{22} \pm \sqrt{1-2\Omega_{21}-2 \Omega_{22} + \Omega_{22}^2-2\Omega_{43} - 2 \sqrt{\left( \Omega_{21}-\Omega_{43} \right)^2 + 4 \Omega_{23} \Omega_{41}}} \right] \;,  \nonumber\\ 
\end{eqnarray} 
where the upper (respectively, lower) sign in front of the external square root refers to $\lambda_1$ and $\lambda_3$ (respectively, $\lambda_2$ and $\lambda_4$). We denote the corresponding eigenmodes by $u^{(i)}$ (we do not report their explicit expressions, as they are lengthy and not illuminating). The eigenvalues and eignvectors are sorted as in   \ref{appA}. Namely, in the limit of two massive  species with $\Omega_A + \Omega_B =1$, and with trivial velocity bias $b_v=1$, the matrix  (\ref{BIGOMEGA}) reduces to (\ref{Om-simple}), and the eigenvalues/eigenmodes $\lambda_i / u^{(i)}$ reduce to (\ref{Om-eigen}), with the same sorting. Specifically, the mode $u^{(1)}$ is the growing mode of the system,  the mode $u^{(2)}$ is the decaying mode present also in the case of a single species, and the modes $u^{(3,4)}$ are the isocurvature modes. We now show that, if the initial conditions are chosen according to the growing mode,  the equal time squeezed bispectrum receives an enhanced contributions only from isocurvature modes internal lines.   

As we already remarked, the term $\propto 1+b_v$ is not enhanced at equal time \cite{Peloso:2013zw}. Therefore, starting from (\ref{treeB}), we have
\begin{eqnarray}
&& \!\!\!\!\!\!\!\!\!\!\!\!\!\!\!\!  \lim_{k\ll q} B^{\vp\vp\vp}_{abc,{\rm tree}} \left( k,q,p;\eta,\eta',\eta' \right) =    \left( 1 - b_v \right) 
  P^{(0)} \left( q \right) P^{(0)} \left( k \right) \frac{\bk \cdot \bq}{2 k^2} \, u_2^{(1)} u_a^{(1)}  \nonumber\\ 
&& \!\!\!\!\!\!\!\!\!\!\!\!\!\!\!\!  \quad\quad\quad\quad  \times \left\{  \int^{\eta'} d s \, {\rm e}^s \; u_b^{(1)} \, g_{cd} \left( \eta' - s ; q \right)  h_{de}   u_e^{(1)} 
    -  \left(     b \leftrightarrow c \right)  \right\}  + {\rm O } \left( k^0 \right) \,. 
\label{squeezed-equaltime}
\end{eqnarray}
where we have remarked that we have chosen initial conditions according to the growing mode. An explicit computation shows that
\begin{equation}
h_{de} u_e^{(1)} = c_1 u_d^{(1)} +  c_3 u_d^{(3)} +  c_4 u_d^{(4)}  \,.
\label{hu1}
\end{equation}
The  expressions for the coefficients $c_{1,3,4}$ can be readily obtained from the explicit computation, but they are not relevant for the present discussion. What is instead relevant for our proof is  that the coefficient $c_2$ vanishes. Thanks to the properties (\ref{prop1}) and (\ref{prop2}) of the linear propagator,we then see that 
\begin{equation}
g_{cd} h_{de} u_e^{(1)} = {\tilde c}_1 u_c^{(1)} +    {\tilde c}_3 u_c^{(3)} +    {\tilde c}_4 u_c^{(4)} \,,   
\label{ghu1}
\end{equation}
where the new coefficients ${\tilde c}_i$ can be readily obtained from (\ref{hu1}) and the explicit form of the linear propagator.  The first term in (\ref{ghu1}) gives a vanishing contribution to  (\ref{squeezed-equaltime}), under the $b\leftrightarrow c$ anti-symmetrization. This verifies at tree level that only internal lines isocurvature modes contribute to the equal time squeezed bispectrum.

\section{Two matter species and $b_v=1$}
  \label{appA}
  
In  \ref{treetest}    we made use of some properties of the eigenmodes of (\ref{BIGOMEGA}). To compare with those discussions, we here briefly outline the properties of the eigenmodes in the most simple case in which there are two matter  species $A$ and $B$, with $\Omega_A + \Omega_B = 1$, and with a trivial velocity bias $b_v = 1$. In this case, the matrix  (\ref{BIGOMEGA}) reads
\begin{equation}
\Omega = \left( \begin{array}{cccc}
1 & - 1 & 0 & 0 \\
- \frac{3 \Omega_A}{2} & \frac{3}{2} & \frac{- 3 \Omega_B}{2} & 0 \\
0 & 0 & 1 & -1 \\
- \frac{3 \Omega_A}{2} & 0 & - \frac{3 \Omega_B}{2} & \frac{3}{2}  
\end{array} \right) \,. 
\label{Om-simple}
\end{equation} 
This matrix has the eigenvalues /  eigenvectors 
\begin{eqnarray}
\lambda_1 = 0 \;\;&,& u^{(1)} = \left( 1 , 1 , 1, 1 \right) \;,  \nonumber\\ 
\lambda_2 = \frac{5}{2} \;\;&,& u^{(2)} = \left( - \frac{2}{3} , 1 , - \frac{2}{3} , 1 \right) \;,  \nonumber\\ 
\lambda_3 = \frac{3}{2} \;\;&,& u^{(3)} = \left( 2 \Omega_B ,\, - \Omega_B ,\, - 2 \Omega_A ,\, \Omega_A \right) \;,  \nonumber\\ 
\lambda_4 = 1 \;\;&,& u^{(4)} = \left(  \Omega_B ,\, 0 ,\,  - \Omega_A ,\, 0  \right) \;.   
\label{Om-eigen}
\end{eqnarray} 
 As clear from (\ref{compact}), in the linear regime the eigenmodes evolve as $u^{(i)} \propto {\rm e}^{-\lambda_i \eta}$. The mode $ u^{(1)} $ is the growing mode of the system (recall that $\delta \propto {\rm e}^\eta u$). The mode $u^{(2)}$ is the decreasing mode of the system that, together with  $ u^{(1)} $, is present also in the case of a single species. The remaining two modes are isocurvature modes, since they satisfy $\Omega_A \delta_A + \Omega_B \delta_B = 0 $. 
 
The linear propagator of the system is given by 
\begin{equation}
g_{ab} \left( \eta \right) = \sum_{i=1}^4 \left( M_i \right)_{ab} {\rm e}^{-\lambda_i \eta} \theta \left( \eta \right) \,, 
\label{prop1}
\end{equation}
where $\theta \left( \eta \right)$ is the  Heaviside theta function, while the projectors $M_i$ are given by
\begin{equation}
\left( M_i \right)_{ab} = \lim_{\lambda \rightarrow \lambda_i} \left( \lambda_i - \lambda \right) \left( \Omega - \lambda
\, 1 \right)^{-1}_{ab} \,. 
\end{equation}
where $1$ denotes the unit matrix. They satisfy
\begin{equation}
\sum_i M_i = 1 \;\;,\;\;   M_i u^{(j)} = \delta_{ij} u^{(i)} \;\;,
\label{prop2}
\end{equation}
as can be also verified from their explicit expressions 
\begin{eqnarray}
&& \!\!\!\!\!\!\!\!\!\!\!\!\!\!\!\! \!\!\!\!\!\!\!\!\!\!\!\!\!\!\!\! \!\!\!\!\!\!
M_1 = \frac{1}{5}  \left( \begin{array}{cccc}
 3 \Omega_A &  2 \Omega_A & 3 \Omega_B & 2 \Omega_B \\
 3 \Omega_A &  2 \Omega_A & 3 \Omega_B & 2 \Omega_B \\
 3 \Omega_A &  2 \Omega_A & 3 \Omega_B & 2 \Omega_B \\
 3 \Omega_A &  2 \Omega_A & 3 \Omega_B & 2 \Omega_B \end{array} \right)\,,\nonumber\\
&&\nonumber\\
&&  \!\!\!\!\!\!\!\!\!\!\!\!\!\!\!\! \!\!\!\!\!\!\!\!\!\!\!\!\!\!\!\! \!\!\!\!\!\!M_2 = \frac{1}{5}  \left( \begin{array}{cccc}
 2 \Omega_A & - 2 \Omega_A & 2 \Omega_B & - 2 \Omega_B \\
- 3 \Omega_A &  3 \Omega_A & -3 \Omega_B & 3 \Omega_B \\
 2 \Omega_A & - 2 \Omega_A & 2 \Omega_B & - 2 \Omega_B \\
- 3 \Omega_A &  3 \Omega_A & -3 \Omega_B & 3 \Omega_B \end{array} \right)\,,\nonumber\\ 
 &&\nonumber\\
&&\!\!\!\!\!\!\!\!\!\!\!\!\!\!\!\! \!\!\!\!\!\!\!\!\!\!\!\!\!\!\!\! \!\!\!\!\!\!
M_3 =   \left( \begin{array}{cccc}
0 &  - 2 \Omega_B &  0 &  2 \Omega_B \\
0 &  \Omega_B &  0 & - \Omega_B \\
0 &  2 \Omega_A &  0 &  - 2 \Omega_A \\
0 &   - \Omega_A &  0 &  \Omega_A \end{array} \right) \;\;,\;\; 
M_4 =   \left( \begin{array}{cccc}
\Omega_B &   2 \Omega_B &  - \Omega_B & - 2 \Omega_B \\
0 &  0 &  0 &  0 \\
- \Omega_A &  - 2 \Omega_A &  \Omega_A &  2 \Omega_A \\
0 &   0 &  0 &  0 \end{array} \right) \;\;. \nonumber\\ 
 \end{eqnarray}

 \section*{References}
\bibliographystyle{JHEP}
\bibliography{/Users/pietroni/Bibliografia/mybib.bib}
\end{document}